\documentclass[aps, pra, twocolumn, superscriptaddress, english]{revtex4}
\usepackage{graphicx}
\usepackage{enumerate}
\usepackage{hyperref}
\usepackage{textcomp}
\usepackage{amsmath}
\usepackage{amssymb}
\usepackage{pgf}
\usepackage{subfigure}
\usepackage[english]{babel}
\usepackage[normalem]{ulem}

\newcommand{\bbeta}{\boldsymbol{\beta}}
\newcommand{\bgamma}{\boldsymbol{\gamma}}

\def\bra#1{\langle#1 |}
\def\ket#1{| #1\rangle}
\newcommand{\braket}[2]{\langle #1 | #2 \rangle}

\newcommand{\nep}{\textrm{e}}

\newcommand{\x}{{\bf x}}
\newcommand{\z}{{\bf z}}

\newcommand{\psitarget}{\psi_{\mathrm{\scriptscriptstyle targ}}}

\newcommand{\calF}{{\mathcal{F}}}

\newcommand{\prm}{\mathrm{p}}

\newcommand{\PauliSigma}{\hat{\sigma}}

\newcommand{\Ho}{\hat{H}}

\newcommand{\Hz}{\widehat{H}_z}
\newcommand{\Hx}{\widehat{H}_x}
\newcommand{\Htarg}{\widehat{H}_{\scriptscriptstyle \mathrm{target}}}
\newcommand{\Spin}{\hat{\mathrm S}}
\newcommand{\Nsites}{\mathrm{N}}
\newcommand{\res}{\mathrm{res}}
\newcommand{\Ptrot}{\mathrm{P}}
\newcommand{\Pcrit}{\mathrm{P}_{\Nsites}^*}

\newcommand{\tann}{\tau}
\newcommand{\pifour}{{\textstyle \frac{\pi}{4}}}
\newcommand{\Ztwo}{{\mathbb Z}_2}
\newcommand{\uvect}{\mathbf{u}}
\newcommand{\vvect}{\mathbf{v}}
\newcommand{\identity}{\hat{1}}
\newcommand{\Nup}{\Nsites^{\uparrow}}
\newcommand{\Ndown}{\Nsites^{\downarrow}}
\newcommand{\Magn}{\mathrm{M}}

\newcommand{\revision}[1]{{\textcolor{black}{#1}}}

\begin{document}

\title{Polynomial scaling of QAOA for ground-state preparation of the fully-connected p-spin ferromagnet}
\author{Matteo M. Wauters}
\affiliation{SISSA, Via Bonomea 265, I-34136 Trieste, Italy}

\author{Glen B. Mbeng}
\affiliation{Universit\"at Innsbruck, Technikerstrasse 21 a, A-6020 Innsbruck, Austria}

\author{Giuseppe E. Santoro}
\affiliation{SISSA, Via Bonomea 265, I-34136 Trieste, Italy}
\affiliation{International Centre for Theoretical Physics (ICTP), P.O.Box 586, I-34014 Trieste, Italy}
\affiliation{CNR-IOM Democritos National Simulation Center, Via Bonomea 265, I-34136 Trieste, Italy}

\begin{abstract}
We show that the quantum approximate optimization algorithm (QAOA) can construct with polynomially scaling resources the ground state of the fully-connected 
p-spin Ising ferromagnet, a problem that notoriously poses severe difficulties to a Quantum Annealing (QA) approach, due to the exponentially small gaps encountered
at first-order phase transition for $\prm\ge 3$. 
\revision{For a target ground state at arbitrary transverse field,} we find that an appropriate QAOA parameter initialization is necessary to achieve a good performance of the algorithm when 
the number of variational parameters $2\Ptrot$ is much smaller than the system size $\Nsites$, because of the large number of sub-optimal local minima.
\revision{Instead, when $\Ptrot$ exceeds a critical value $\Pcrit \propto \Nsites$,} the structure of the parameter space simplifies, as all minima become degenerate.
This allows to achieve the ground state with perfect fidelity with a number of parameters scaling extensively with $\Nsites$, and with resources scaling polynomially with $\Nsites$.
\end{abstract}

\maketitle
\section{Introduction} 
Efficient optimization and ground state (GS) preparation are two of the most prominent issues in the growing field of quantum technology~\cite{QuantumTech_2018,Kurizki_PNAS15}.
Optimization is a long-standing problem in physics and computer science, and lies at the roots of the efforts to show a possible ``quantum supremacy'' \cite{Preskill_arXiv2012, Farhi_arXiv2016,Nature_QuantumSupremacy} over classical algorithms.
A robust state preparation strategy, in turn, would be a crucial tool for quantum technologies, and would also allow to ``solve'', using quantum hardware, many long-standing problems
in condensed matter theory or quantum chemistry~\cite{Lloyd_SCI96,Buluta_SCI09,Nori_RMP14}. 
The two are intimately connected, as many optimization tasks can be reformulated in terms of finding the classical ground state of an appropriate spin-glass 
Hamiltonian~\cite{Lucas_FrontPhys2014}.

A traditional tool in this field has been Quantum Annealing~\cite{Finnila_CPL94,Kadowaki_PRE98,Brooke_SCI99,Santoro_SCI02} (QA) 
--- {\em alias} Adiabatic Quantum Computation~\cite{Farhi_SCI01,Albash_RMP18} ---,
which relies on the adiabatic theorem to find the ground state of a target Hamiltonian, starting from a trivial initial state.
Although QA appeared to be more efficient than its classical counterpart for certain problems~\cite{Santoro_SCI02,Heim_SCI15,Crosson_FOCS2016,Naven_PRX2016,Albash_PRX2018}, 
it is limited by the smallest gap encountered during the evolution, which vanishes, in the thermodynamic limit, when the system crosses a phase transition.
In this context, the fully-connected p-spin Ising ferromagnet is a simple but useful benchmark for optimization methods, because QA fails due to the exponentially small gap
at the first-order phase transition encountered for $p\ge 3$~\cite{Jorg_EPL10,Bapst_JSTAT12, Wauters_PRA17}.
Several techniques have been advocated to overcome the slowness induced by such an exponentially small gap, including the introduction of non-stoquastic terms~\cite{Nishimori_ICT2017,Nishimori_JPSJ18}, pausing~\cite{Passarelli_PRB19}, dissipative effects~\cite{Passarelli_PRA19,Passarelli_PRA20}, or approximated counterdiabatic driving~\cite{Passarelli_PRR20}.
\revision{Their successful application, however, often depends on the knowledge of the spectrum or on the phase diagram of the model, thus making these techniques highly problem-specific. }

Recent alternative ground state preparation approaches~\cite{Peruzzo_NatComm14,Farhi_arXiv2014,Zoller_NAT19} rely on hybrid quantum-classical variational techniques~\cite{McClean_NewJPhys2016} to tackle such problems, avoiding the limitations imposed by a QA adiabatic evolution. 
In this work, we will focus on one such scheme, the Quantum Approximate Optimization Algorithm (QAOA)~\cite{Farhi_arXiv2014, Niu_arx2019, Zhou_PRX2020}.

The core idea of QAOA is to write a trial wavefunction as a product of many unitary operators, each depending on a classical variational parameter, applied to a state simple to construct, usually a product state with spins aligned in the $x$-direction.  
A quantum hardware performs the discrete quantum dynamics and measures the expectation value of the target Hamiltonian, which is then minimized by an external classical algorithm,
as a real function in a high dimensional parameter space.

Although QAOA is a universal computational scheme~\cite{Lloyd_arXiv2018}, its performance strongly depends on the details of the target Hamiltonian. 
QAOA seems to perform rather well on Max-Cut problems~\cite{Zhou_PRX2020} and on short-range spin systems~\cite{Wang_PRA2018, Mbeng_arx19}.
The Grover search problem has also been studied within QAOA, showing that it leads to the optimal square root speed-up with respect to classical algorithms~\cite{Zhang_PRA2017}.   
For generic long-ranged Hamiltonians, however, many open questions remain. 
The questions concern, in particular, the efficiency of the algorithm when a large number of unitaries are employed, or the ability to deal with first-order phase transitions, or the existence of ``smooth'' sets of optimal parameters \cite{Zhou_PRX2020,Mbeng_arXiv2019_digitizedQC,Pagano_arXiv2019}.
Addressing these issues, an essential step towards experimental implementations of QAOA in realistic problems, will be the goal of our work.
We will show that QAOA can construct with polynomially scaling resources the ground state of the fully-connected 
p-spin Ising ferromagnet for all $\prm\ge 2$, hence including the case where a first-order phase transition occurs.
For a generic target state, we find that an appropriate QAOA parameter initialization is necessary to achieve a good performance of the algorithm when 
the number of variational parameters $2\Ptrot$ is much smaller than the system size $\Nsites$, because of the large number of sub-optimal local minima.
Finally, we show that when $\Ptrot > \Pcrit \propto \Nsites$, the structure of the parameter space simplifies, and all minima become degenerate.
This allows to achieve the ground state with perfect fidelity with a number of parameters scaling extensively with $\Nsites$, and with resources scaling polynomially with $\Nsites$.

\revision{The rest of the paper is organized as follows:
in sec.~\ref{sec:model} we present the model and describe the QAOA algorithm.
In sec.~\ref{sec:results} we present our main analytical and numerical results; the technical details of the analytical proof are reported in the appendix \ref{app:P1}. 
Finally, we draw our conclusions and discuss on future outlooks in sec.~\ref{sec:conclusion}.}

\section{Model and QAOA algorithm}\label{sec:model}
As a benchmark for QAOA on long-range models we focus on the ferromagnetic 
fully-connected $\prm$-spin model~\cite{Jorg_EPL10,Vidal_PRA11,Bapst_JSTAT12,Caneva_PRB07,Wauters_PRA17}:
\begin{equation} \label{eqn:model}
\Htarg = -\frac{1}{\Nsites^{\prm-1}} \Big( \sum_{j=1}^{\Nsites} \PauliSigma^{z}_j \Big)^{\prm} - h \Big( \sum_{j=1}^{\Nsites} \PauliSigma^{x}_j \Big) \;,
\end{equation}
where $\PauliSigma^{x,z}_j$ are Pauli matrices at site $j$, $\Nsites$ is the total number of sites, and $h$ a transverse field.
This model displays, for $\prm=2$, a second-order quantum phase transition, at a critical transverse field $h_c=2$, from a paramagnetic ($h>h_c$) to a symmetry-broken 
ferromagnetic phase ($h<h_c$). 
The transition becomes first-order for $\prm>2$, and $h_c$ decreases for increasing $\prm$, with $h_c\to 1$ for $\prm \to \infty$ \cite{Bapst_JSTAT12}.

The QAOA algorithm~\cite{Farhi_arXiv2014} is a variational method to find the ground state of a target Hamiltonian $\Htarg$.
Starting from an initial spin state polarized along the $\hat{\x}$ direction 
$|+\rangle=2^{-N/2}\left( |\!\uparrow\rangle + |\!\downarrow\rangle \right)^{\otimes N}$, 
QAOA writes the following variational {\em Ansatz}
\begin{equation} \label{eq:QAOA1}
|\psi_{\Ptrot}(\bgamma,\bbeta )\rangle 
= \nep^{-i \beta_{\Ptrot} \Hx } \nep^{-i \gamma_{\Ptrot} \Hz} \cdots \nep^{-i \beta_1 \Hx } \nep^{-i \gamma_1 \Hz} | +\rangle
\end{equation}
%
in terms of $2\Ptrot$ variational parameters $\bgamma=(\gamma_1 \cdots \gamma_{\Ptrot})$ and $\bbeta=(\beta_1 \cdots \beta_{\Ptrot})$,
where $\Hz$ and $\Hx$ are non-commuting Hamiltonians depending on the problem we wish to solve.
Here we take $\Hx=-\sum_j \PauliSigma^x_j$, the standard transverse field term, and an interaction term $\Hz$ 
\begin{equation}
\Hz=-\Big( \sum_{j=1}^{\Nsites} \PauliSigma^{z}_j \Big)^{\prm} \;,
\end{equation} 
chosen for convenience to have a super-extensive form with an integer spectrum. 
These choices allow us to restrict the parameter space for $\gamma_m$ and $\beta_m$ to the interval $[0,\pi]$.
%
%
In each QAOA run the variational energy cost function
\begin{equation}\label{eq:QAOA_energy}
E_{\Ptrot}(\bgamma,\bbeta )=\langle \psi_{\Ptrot}(\bgamma,\bbeta)| \Htarg |\psi_{\Ptrot}(\bgamma,\bbeta) \rangle \;,
\end{equation}
is minimized, until convergence to a local minimum $(\bgamma^*,\bbeta^*)$ is obtained.
The quality of the variational solution is gauged by computing the residual energy density~\cite{Mbeng_arx19}  
\begin{equation}\label{eq:rese}
\epsilon_{\Ptrot}^{\res} (\bgamma^*,\bbeta^* ) = \frac{E_{\Ptrot}(\bgamma^*,\bbeta^* )-E_{\min}}{E_{\max}-E_{\min}} \;,
\end{equation}
where $E_{\min}$ and $E_{\max}$ are the lowest and largest eigenvalues, respectively, of the target Hamiltonian.

The connection with a QA approach is interesting~\cite{Mbeng_arx19}. In QA one would write an interpolating Hamiltonian~\cite{Albash_RMP18} 
$\Ho(s)= s \Htarg + (1-s) \Hx$, with $s(t)$ driven from $s(0)=0$ to $s(\tann)=1$ in a sufficiently large annealing time $\tann$. 
A lowest-order Trotter decomposition of the corresponding step-discretized evolution operator 
--- with $s_{m=1\cdots \Ptrot}$ constant for a time-interval $\Delta t_{m=1\cdots \Ptrot}$ ---
would then result in a state of the form of Eq.~\eqref{eq:QAOA1} with:
\begin{equation}\label{eq:l-init}
\left\{
\begin{array}{ll}
\displaystyle \gamma_m = \frac{s_m \Delta t_m}{\hbar} \frac{1}{\Nsites^{\prm-1}} \vspace{3mm}\\
\displaystyle \beta_m = \frac{\Delta t_m}{\hbar} \Big(1 - s_m (1-h)\Big)
\end{array}
\right.
\end{equation}
where the total evolution time would be given by:
\begin{equation} \label{eqn:tau} 
\frac{\tann}{\hbar} = \sum_{m=1}^{\Ptrot} \frac{\Delta t_m}{\hbar} =  \sum_{m=1}^{\Ptrot} \Big( \beta_m + (1-h) \gamma_m \Nsites^{\prm-1} \Big) \;.
\end{equation}
While an optimization of the parameters $s_m$ and $\Delta t_m$ is in principle possible, the standard linear schedule $s(t)=t/\tann$ would result in
a digitized-QA scheme where $s_m=m/\Ptrot$ and $\Delta t_m=\Delta t=\tann/\Ptrot$~\cite{Martinis_Nat16,Mbeng_dQA_PRB2019}. 
With these choices, a convenient starting point for the QAOA optimization algorithm would be to take
$\gamma_m^0 = \frac{\Delta t}{\hbar} \frac{m}{\Ptrot} \frac{1}{\Nsites^{\prm-1}}$ and $\beta_m^0 = \frac{\Delta t}{\hbar} \big(1 - \frac{m}{\Ptrot} (1-h)\big)$
%
%
with possible addition of a small noise term. 
Alternatively, we might choose a completely random initial point with $\gamma_m^0, \beta_m^0 \in [0,\pi]$.
\revision{Limiting the variational parameters in the interval $[0,\pi]$ is motivated by the symmetries of the function $E_{\Ptrot}(\bgamma,\bbeta )$, which we describe in appendix \ref{app:sym}.}
These two alternative choices will be henceforth referred to as l-init and r-init. 


\section{Results}\label{sec:results}
Ref.~\cite{Ho_PRA19} has shown that the target ground state of the $\prm=2$ fully connected Ising ferromagnet with $h=0$, the so-called 
Lipkin-Meshov-Glick~\cite{Lipkin_NucPhys65} model, can be perfectly constructed, with unit fidelity, with the shortest QAOA circuit, $\Ptrot=1$, \revision{if the number of sites $\Nsites$ is odd. For $\Nsites$ even instead, $\Ptrot=2$ is required to reach exactly the GS.}
Ref.~\cite{Streif_arXiv19} has recently shown that a whole class of spin-glass models can be constructed where QAOA shows such a property.
Here we show --- see detailed proof in appendix \ref{app:P1} --- that the general $\prm$-spin model in Eq.~\eqref{eqn:model} belongs, 
for $h=0$, to the class of $\Ptrot=1$ QAOA-solvable problems, for $\Nsites$ odd. 
\revision{The proof is based on finding a set of sufficient conditions to have unit fidelity $\calF(\gamma,\beta)=\Big| \braket{\psitarget}{\psi_{\Ptrot=1}(\gamma,\beta )}\Big|^2$. This provides a set of parameters $(\gamma,\beta)$ that can be used to prepare the exact ground state for $h=0$.} 
For $\Ptrot=1$ the target state fidelity reads: 
\begin{eqnarray}\label{eq:fidelity}
\calF(\gamma,\beta) &=& 
\left| \bra{\psitarget} \nep^{-i\beta \Ho_x} \nep^{-i \gamma \Ho_z}\ket{+}\right|^2  \nonumber \\ 
&=& \Big| \frac{1}{\sqrt{2^\Nsites}} \sum_l \nep^{-i\gamma E_l} \bra{\psitarget} \nep^{-i\beta \Ho_x} \ket{l} \Big|^2 \;,
\end{eqnarray}
where $\ket{\psitarget}$ is the $h=0$ target ground state, and the sum in the second line runs over the $2^{\Nsites}$ basis states $\ket{l}$ of the computational basis, 
with $\Ho_z \ket{l} =E_l \ket{l}$.
Eq.~\eqref{eq:fidelity} shows that $\calF$ is the scalar product of two $2^{\Nsites}$-dimensional unit vectors of components
\begin{equation}
\left\{
\begin{array}{ll}
({\bf v}(\gamma))_l = \nep^{i \gamma E_l}/\sqrt{2^{\Nsites}}\ , \\
({\bf u}(\beta))_l= \langle \psitarget | \nep^{-i\beta \Ho_x} \ket{l}\ ,
\end{array}
\right.
\end{equation}
%
i.e., $\calF = | {\bf v}^{\dagger} \cdot {\bf u} |^2$. 
To ensure $\calF=1$, the Cauchy-Schwarz inequality requires ${\bf v}(\gamma)$ and ${\bf u}(\beta)$ to be parallel up to an overall phase factor. 
As discussed in appendix \ref{app:P1}, this requires $\beta=\frac{\pi}{4}$. 
A unit fidelity further imposes~\cite{Streif_arXiv19} that all terms appearing in the sum in Eq.~\eqref{eq:QAOA1} are pure phase factors, which have to be identical for all $l$, modulo $2\pi$.
In appendix \ref{app:P1} we perform the calculation explicitly, showing that \revision{the pair $\left(\beta=\frac{\pi}{4}, \gamma = \frac{\pi}{4}\right)$ attains unit fidelity} $\calF=1$ for $\prm$ odd, while for $\prm$ even the precise value of $\gamma$ depends
on $\prm$.
\revision{As a remark, notice that in the theoretical proof we use for convenience the fidelity, instead of the residual energy.
In general, however, we prefer the latter as a figure of merit, since it is directly linked to the variational minimization the expectation value of the target Hamiltonian.
Moreover computing the fidelity requires the full knowledge of the target ground state, which in general is not available for large systems.
The energy instead is computed more easily and it is accessible also in experimental implementation of QAOA~\cite{Pagano_arXiv2019}, without performing full tomography of the variational state.}

The possibility of preparing exactly the GS with $\Ptrot=1$ is noteworthy, as it suggests that one can construct the exact $h=0$ classical ground state with 
an algorithm whose equivalent computational time, see Eq.~\eqref{eqn:tau}, scales as $\Nsites^{\prm-1}$. 
On the contrary, for any finite $\Nsites$, a QA algorithm would need to cope with a minimum spectral gap at the transition
point~\cite{Jorg_EPL10,Bapst_JSTAT12,Caneva_PRB07,Wauters_PRA17,Defenu_PRL18} which scales as $\Delta \sim \Nsites^{-1/3}$ if $\prm=2$ and 
$\Delta\sim\nep^{-\alpha_\prm \Nsites}$ if $\prm \geq 3$: 
with a linear schedule annealing, this implies a total annealing time $\tann\propto \Delta^{-2}$, hence 
$\tau \sim \Nsites^{2/3}$ for $\prm=2$ and $\tau \sim \nep^{2\alpha_\prm \Nsites}$ for $\prm>2$. 
Therefore, QAOA shows an {\em exponential speed-up} with respect to a linear-schedule QA for $\prm>2$, \revision{without exploiting any knowledge on the spectrum or on the phase diagram.}

\begin{figure}
\begin{center}
\includegraphics[width=8cm]{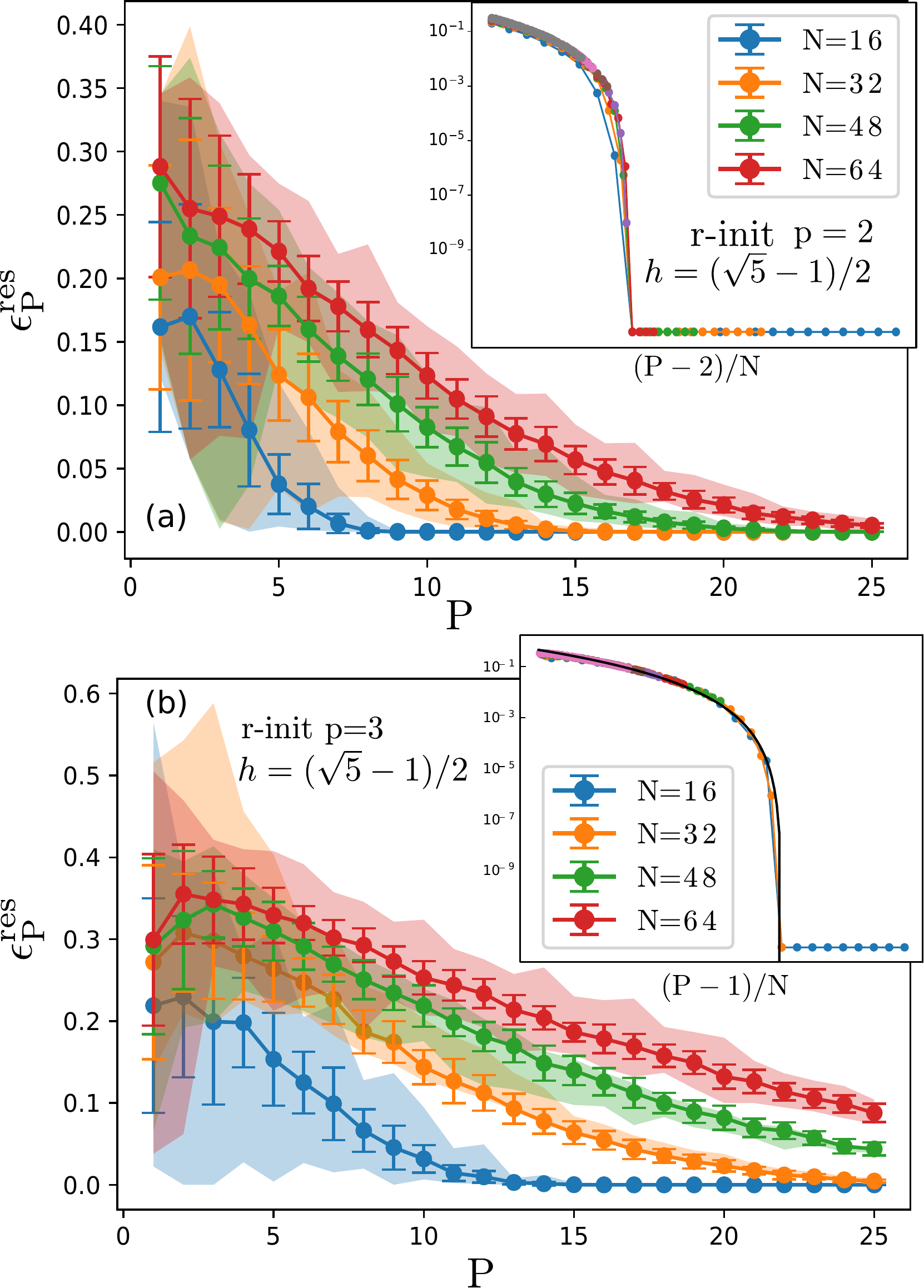}
\caption{(a) Results of local optimizations with $(\bgamma^0,\bbeta^0)$ initialized randomly in $[0,\pi]$ (r-init) averaged over 100 different realizations, 
for several values of $\Nsites$ and $h=\frac{\sqrt{5}-1}{2}$, for $\prm=2$.   
\revision{The shaded areas show the range between the best and the worst result obtained for each set of data.}
The inset shows the collapsed data (in log scale) after rescaling $\Ptrot \to (\Ptrot-2)/\Nsites$.
(b) same data for $\prm=3$. The rescaling in the inset now is $\Ptrot \to (\Ptrot-1)/\Nsites$ and the black line is the curve $\bigg(1-\frac{\Ptrot}{\Pcrit} \bigg)^3$.
Results for larger $\prm>3$ are qualitatively similar.}
\label{fig:localOpt_p2}
\end{center}
\end{figure}
Such a remarkable property is however lost as soon as one targets a ground state with $h\neq 0$, 
where QAOA is no longer able to find the exact ground state with a \revision{small parameter number}, $\Ptrot=1$ or $2$.
We find that the energy landscape $E_{\Ptrot}(\bgamma,\bbeta)$ is extremely rugged for $\Ptrot>2$, making local optimizations 
--- specifically, we use the Broyden-Fletcher-Goldfard-Shanno (BFGS) algorithm~\cite{Nocedal_book2006} ---
highly dependent on the initial set of parameters $(\bgamma^0,\bbeta^0)$.
We observe a very different behavior if the minimization is initialized with parameters $\gamma_m^0$ and $\beta_m^0$ chosen randomly in $[0,\pi]$ (r-init), 
or rather with an initial guess based on a linear schedule,
$\gamma_m^0 = \frac{\Delta t}{\hbar} \frac{m}{\Ptrot} \frac{1}{\Nsites^{\prm-1}}$ and $\beta_m^0 = \frac{\Delta t}{\hbar} \big(1 - \frac{m}{\Ptrot} (1-h)\big)$ (l-init).
The results for the random initialization are summarized in Fig.~\ref{fig:localOpt_p2}, where we show the normalized residual energy, Eq.~\eqref{eq:rese}, 
versus the number of QAOA steps $\Ptrot$ for $h=\frac{\sqrt{5}-1}{2}<h_c$, whose target state lies in the ferromagnetic phase for any value of $\prm$.
Data for different system sizes $\Nsites$ collapse perfectly after rescaling $\Ptrot \to (\Ptrot-2)/\Nsites$ (see inset of Fig.~\ref{fig:localOpt_p2}(a)) and 
drop below machine precision at $\Ptrot = \Pcrit =\frac{\Nsites}{2}+2$.
Correspondingly, the variance of the residual energy distribution, which is rather large for $\Ptrot < \Pcrit$ as witnessed by the error bars, 
drops to $0$ at $\Pcrit$, implying that all local minima become degenerate. 
\revision{The colored area around each curve shows the range between the lowest and the highest residual energy obtained for each value of $\Ptrot$ and $N$.
The distribution of individual optimizations is symmetric around the average, with the exception of small values of $\Ptrot$ where r-init QAOA occasionally finds a local minimum with very small residual energy.
For a better readability, in the following figures we report only errorbars corresponding to the standard deviation of our data.}

The scaling shown in fig.~\ref{fig:localOpt_p2} holds for any value of the transverse field $h$, if the QAOA minimization is initialized with random parameters. 
In general, we find that the residual energy follows:
\begin{equation}\label{eq:random_eres}
\epsilon^{\res}_{\Ptrot}= \left\{
\begin{array}{ll}
\displaystyle \bigg(1-\frac{\Ptrot}{\Pcrit} \bigg)^b \; & \mbox{if  } \; \Ptrot < \Pcrit \vspace{3mm} \\
\displaystyle0 \;  & \mbox{if  } \;   \Ptrot \ge \Pcrit 
\end{array}
\right. \;,
\end{equation}
with $b\simeq 3$.
Remarkably, this scaling holds also for $\prm>2$, with similar values of $b$, 
with the only difference that $\Pcrit = \Nsites+1$ for $\prm$ odd, because of the lack of the $\Ztwo$ symmetry.
This in turn implies that for {\em finite} $\Nsites$ one can attain a perfect control of the state with $\Ptrot=\Pcrit \propto \Nsites$, 
physically corresponding to a total evolution time that scales as a power-law with $\Nsites$. 
\revision{Our data for $\prm=3$ are reported in Fig.~\ref{fig:localOpt_p2}(b), where, in the inset, we also highlight the curve describe by Eq.~\eqref{eq:random_eres} (solid black line).}
Once again, this is at variance with a standard linear-schedule QA, where the total evolution time has to scale
{\em exponentially} with $\Nsites$ when the transition is first order, i.e., for $\prm>2$. 

\begin{figure}
\begin{center}
\includegraphics[scale=0.55]{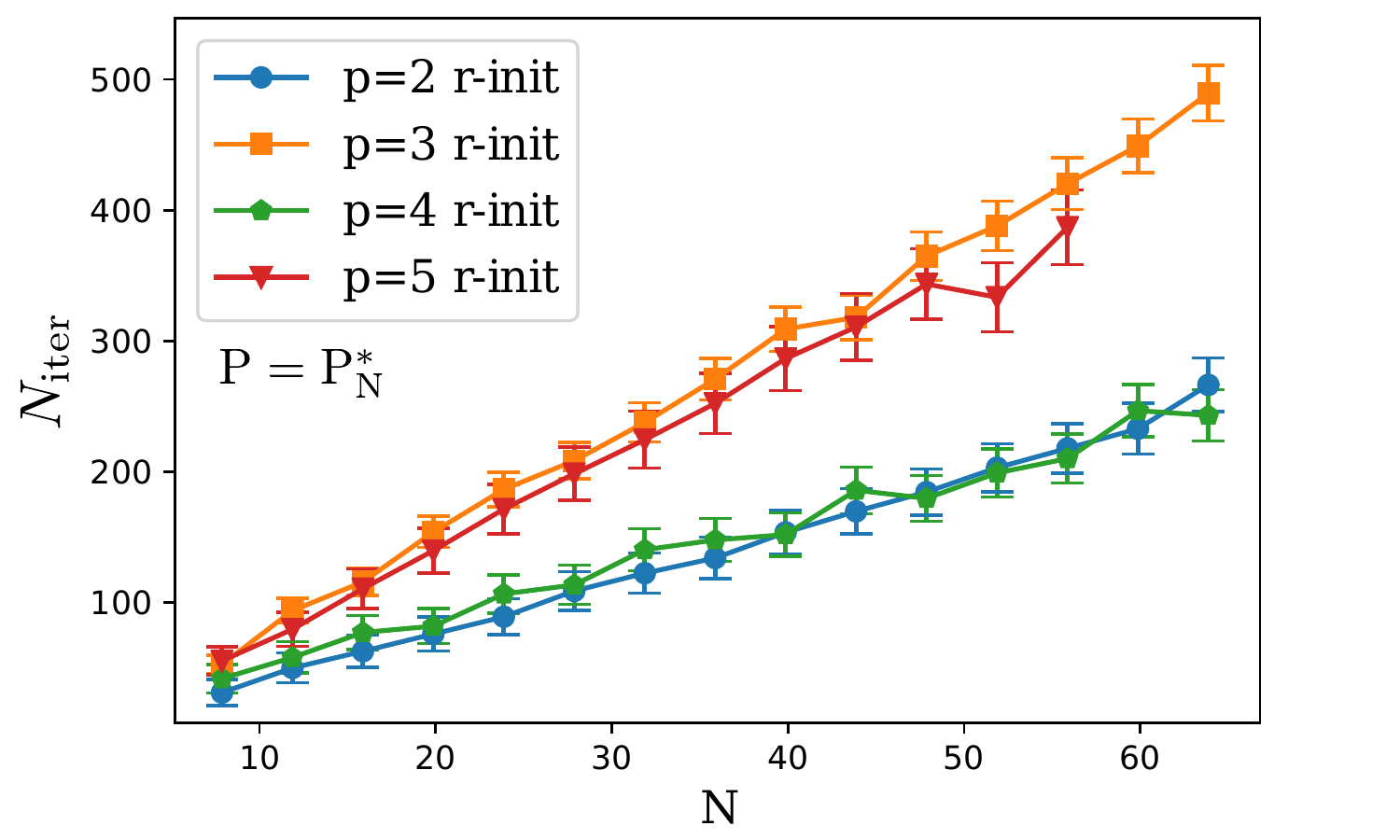}
\caption{Number of iterations required for BFGS algorithm to converge, averaged over 20 optimizations with random initializations of QAOA parameters. 
The corresponding value of $\Ptrot$ is $\Nsites/2+2$ for $\prm$ even, and $\Nsites+1$ for $\prm$ odd, which is sufficient to obtain a residual energy below the numerical error.}
\label{fig:Pcrit_scaling}
\end{center}
\end{figure}
\revision{
    We have shown that a QAOA circuit with $\Ptrot=\Pcrit \propto \Nsites$ is sufficient to prepare the exact ground state of the $\prm$-spin model for an arbitrary target $h$. However, to estimate the total computational complexity of running the QAOA algorithm to solve the $\prm$-spin model, we must include the computational cost of finding the QAOA variational parameters (with BFGS). Indeed, during the optimization process, the quantum device is used $N_{\mathrm{iter}}$ times, to sample the optimization landscape associated with QAOA circuits of $\Ptrot=\Pcrit$. In Fig.~\ref{fig:Pcrit_scaling} we show the number of iterations $N_{\mathrm{iter}}$, that the BFGS required for convergence as a function of $N$.
    $N_{\mathrm{iter}}$ appears to increase linearly with $N$, with an angular coefficient that only depends on the parity of $\prm$. Hence,  the total computational time needed for converging to the exact ground state, at arbitrary transverse field, is at most polynomial in $N$,
    since it requires an order $O(N)$ of iterations and a similar number of variational parameters, all in the range $[0,\pi]$. }

\begin{figure}
\begin{center}
\includegraphics[scale=0.55]{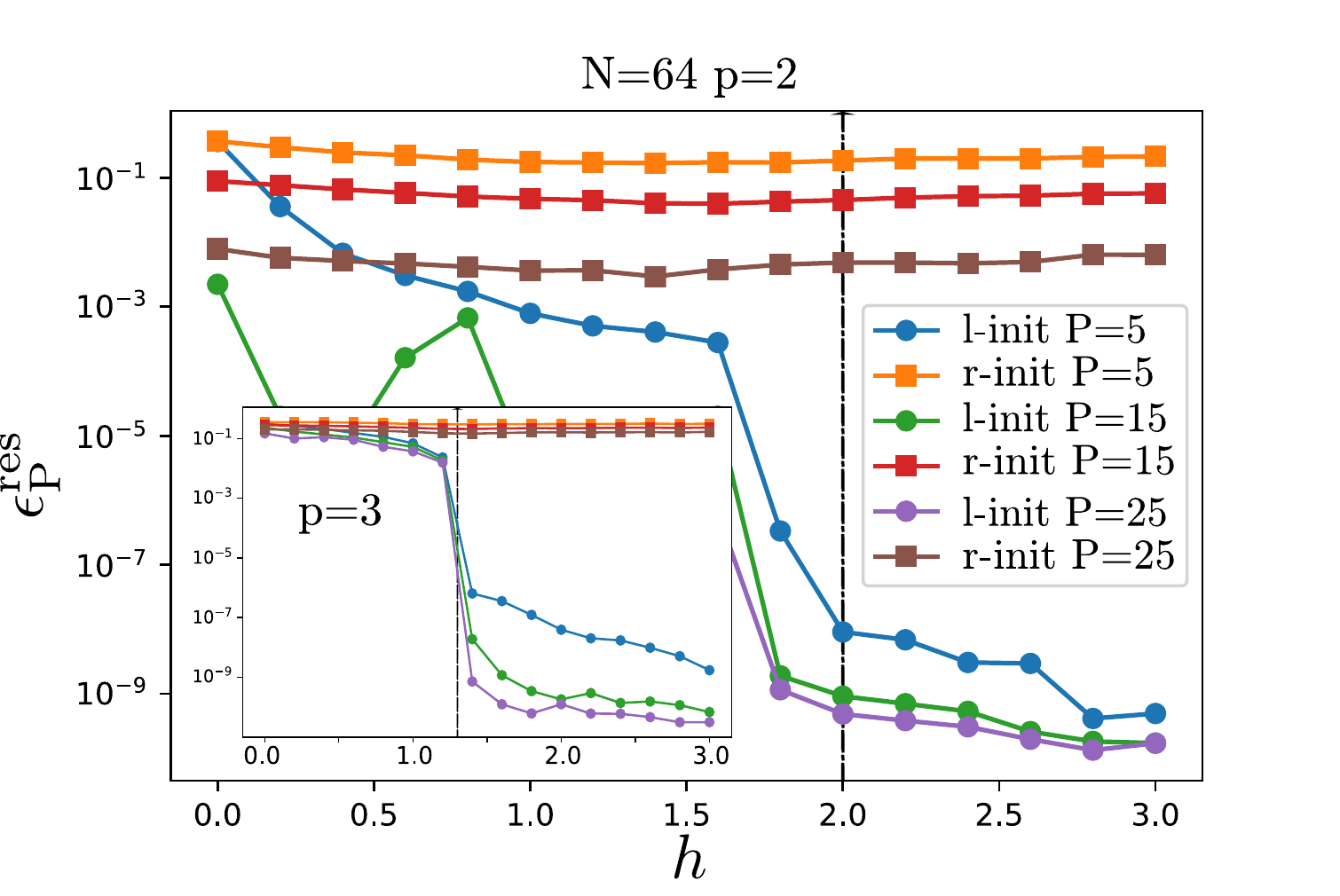}
\end{center}
\caption{QAOA residual energy versus the transverse field $h$ for a system with $\Nsites=64$, and three values of $\Ptrot = 5, 15, 25$,
for random (r-init) and linear initialization (l-init) of the QAOA parameters. 
Notice that $\Pcrit=\Nsites/2+2=34$ for $\prm=2$ and $\Pcrit=\Nsites+1=65$ for $\prm=3$. 
The vertical dashed lines show the critical transverse field: $h_c=2$ for $\prm=2$ (main plot) and $h_c\simeq 1.3$ for $\prm=3$ (inset).
The linear initialization (l-init) corresponds to Eq.~\eqref{eq:l-init} multiplied element-wise by a noise factor $(1+r)$,
 with $r \in [-0.05,0.05]^{2\Ptrot}$ a vector of uniformly distributed random numbers. 
Data are averaged over 100 different instances of $r$.
}
\label{fig:p2_opt_lin_g}
\end{figure}

A linear initialization of QAOA parameters, with a small noise (see caption of Fig.~\ref{fig:p2_opt_lin_g} for details), improves drastically the QAOA performance. 
This is illustrated in Fig.~\ref{fig:p2_opt_lin_g} where the results of the two competing schemes, random (r-init) versus linear (l-init) initialization, are
shown for a system with $\Nsites=64$ for both $\prm=2$ (main plot) and $\prm=3$ (inset), and three fixed values of $\Ptrot=5, 15, 25$.  
Notice how the linear initialization is able to ``detect'' the quantum paramagnetic phase, for $h>h_c$, as being ``easy'', with 
the QAOA minima found having vanishingly small residual energy, almost to machine precision, even if $\Ptrot < \Pcrit$.
This occurs not only in the second-order transition case with $\prm=2$, but also in the more ``difficult'' first-order case with $\prm=3$.
At variance with that, a random initialization performs on average quite independently of the target transverse field $h$, and knows nothing about the location of
the critical field.
\revision{Interestingly, this suggests that QAOA is sensitive to the phase diagram of the target Hamiltonian: choosing a good ansatz for the initial parameter set $(\bgamma^0,\bbeta^0)$ is fundamental to initialize the variational wavefunction in a ``good'' basin of attraction, possibly in the same phase of the target state, where the minimization leads to small values of $\epsilon^\res_\Ptrot$.
Whether this feature is unique to infinite range models or is a common property of long range Hamiltonians is an interesting issue to pursue in future works.}

\begin{figure}
\begin{center}
\includegraphics[scale=0.5]{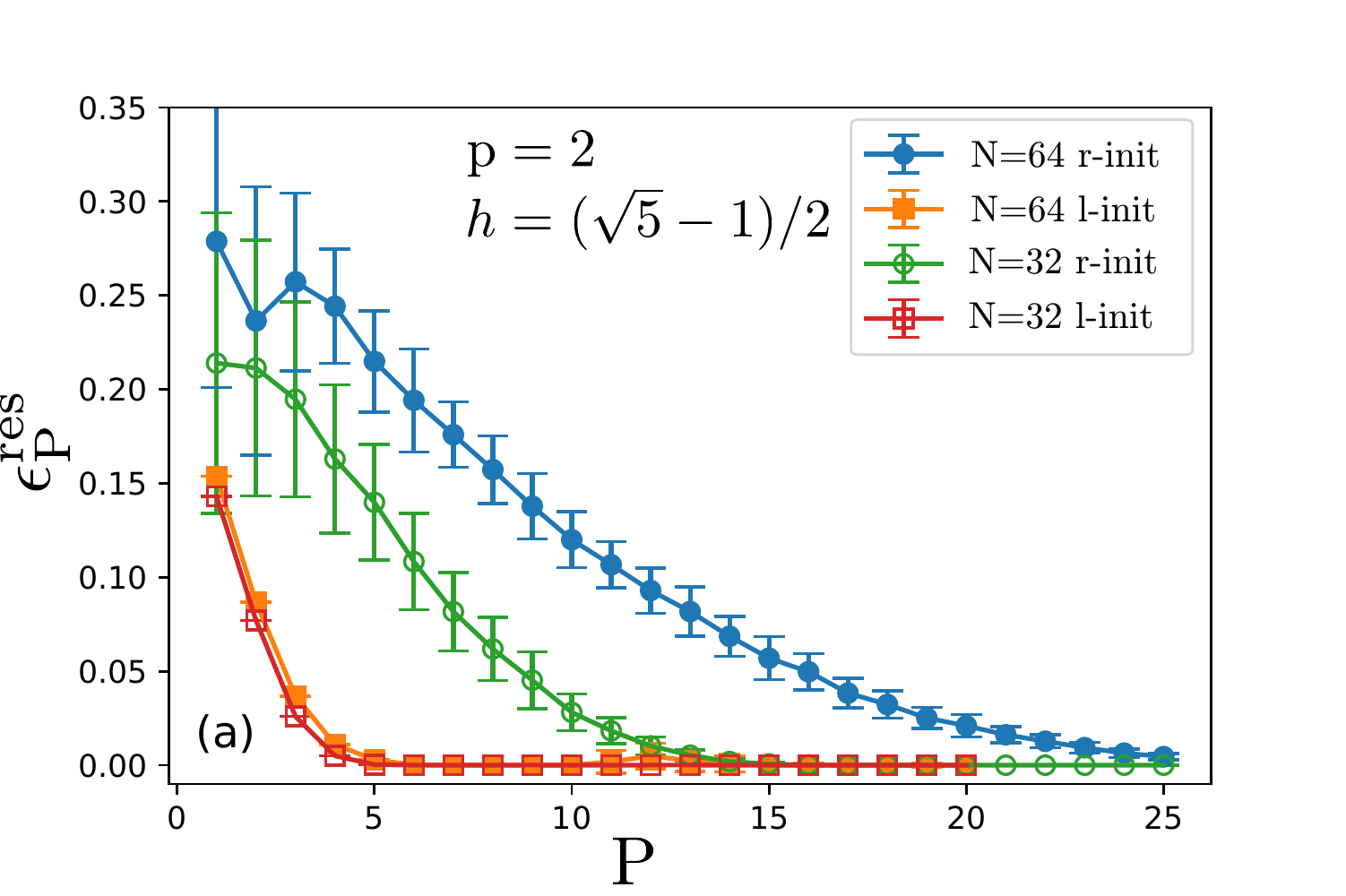}
\includegraphics[scale=0.5]{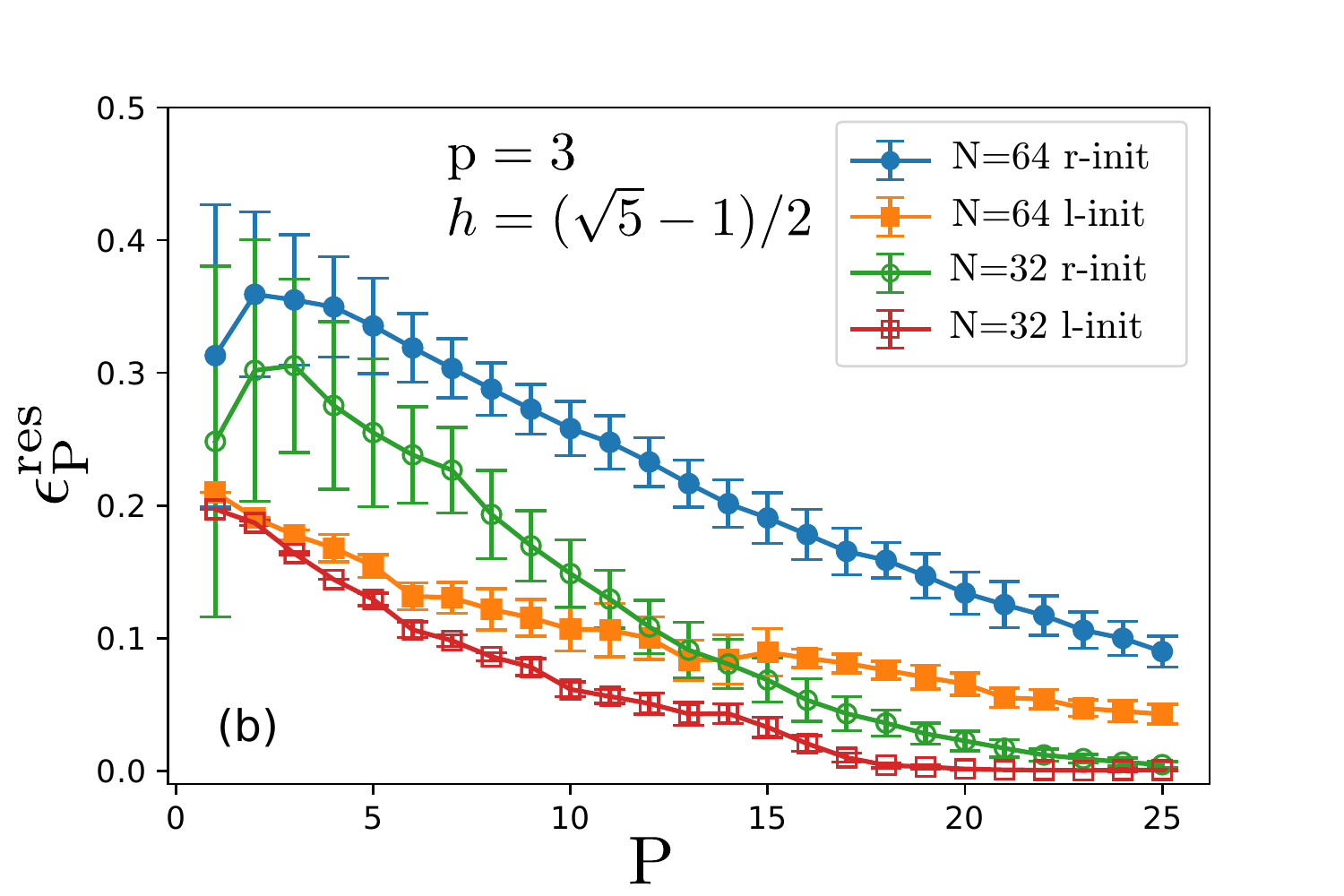}
\end{center}
\caption{Comparison between the optimized residual energy obtained from a linear initial guess plus small noise (l-init) and from random initialization (r-init), 
for two system sizes $\Nsites=32$ and $\Nsites=64$. In (a) $\prm=2$, in (b) $\prm=3$.}
\label{fig:p2_opt_lin}
\end{figure}

The linear initialization displays better efficiency, compared to the random one, also when the target state belongs to the ferromagnetic phase 
($h<h_c$), and $\Ptrot < \Pcrit$. This is illustrated in Fig.~\ref{fig:p2_opt_lin} for $\prm=2$ (a) and $\prm=3$ (b). 
Here, however, the improvement is only quantitative --- $\epsilon^{\res}_\Ptrot$ decreases faster and scales better
with system size --- since the actual change in the landscape, with degenerate global minima, occurs only at $\Pcrit$.
Moreover, the system displays a large roughness of the variational energy landscape, which makes
the task of obtaining good variational minima extremely demanding, especially for $\prm\ge 3$, 
hence justifying the poorer improvement of l-init over r-init observed in Fig.~\ref{fig:p2_opt_lin}(b). 

\revision{A smooth change of the control parameters is required, or at least useful, for experimental implementations of QAOA algorithms~\cite{Pagano_arXiv2019}.
Beside allowing for an easier control of external fields, they also leads to faster convergence to a local minimum~\cite{Mbeng_arx19}, hence reducing the number of measurements to be performed.
Finding local minima $(\bgamma^*,\bbeta^*)$ which can be seen as the discretization of some continuous function, proves however to be a difficult task for this model.
In contrast with Refs.~\cite{Zhou_PRX2020,Mbeng_arx19}, an iterative procedure that initializes $(\bgamma^0,\bbeta^0)$ from an interpolation of a smooth set 
obtained for a smaller parameter space does not seem to work in a straightforward way.
The linear initializazion we have adopted is able to find reasonably smooth $(\bgamma^*,\bbeta^*)$ only for small values of $\Ptrot$. 
As the dimensionality of the parameter space increases, and so does the roughness and the number of local minima, the optimal parameters obtained starting 
from a linear initialization scheme appear to be increasingly irregular (data not shown).
Our failed attempts do not exclude that smart smooth choices for $(\bgamma^0,\bbeta^0)$ can be constructed: they only signal that finding them is a non-trivial task, due 
to the extreme roughness of the energy landscape.
Recent results of Reinforcement Learning assisted QAOA show that smooth protocols can be constructed~\cite{Wauters_RL_arx2020}; if they describe the discretization of an adiabatic schedule or not is still an open question.}

\section{Conclusions}\label{sec:conclusion}
We analyzed the performance of QAOA on the fully connected p-spin model, showing that it is able to find exactly the ferromagnetic ground state with polynomial resources, even when the system encounters a first order phase transition. 
In particular, the algorithm prepares the ground state of $\Hz$ with only $\Ptrot=1$ (if $\Nsites$ is odd) or $\Ptrot=2$ (if $\Nsites$ is even) steps, 
with a corresponding evolution time that scales as $\Nsites^{\prm-1}$, while QA would require an exponentially long annealing time.
This exact minimum however exists only for zero transverse field, $h=0$. 
Interestingly, the exact minimum, which clearly survives for $\Ptrot\geq 2$, is very hard to find with gradient-based optimization schemes due to the extreme roughness 
of the energy landscape, especially for $\prm>2$.  
The ``hardness'' of the problem for $\prm >2$ is thus reflected in the difficulty in finding the correct absolute minimum, rather than in the resources (i.e. the computational time) needed.

The performance of the optimization itself strongly depends on the initialization of the variational parameters $(\bgamma^0,\bbeta^0)$.
For a random initialization, the residual energy drops below machine precision as $(\Pcrit-\Ptrot)^b$, with $b\sim 3$ and $\Pcrit$ growing linearly with $\Nsites$.
This behavior is independent of the target transverse field $h$ and from $\prm$, with the only difference that $\Pcrit=\Nsites/2+2$ for $\prm$ even and 
$\Pcrit=\Nsites+1$ for $\prm$ odd.
With a linear initialization, the algorithm performs much better and is able to detect the presence of a phase transition, although the improvement deteriorates 
rapidly as $\Ptrot$ increases, because of the growing number of ``bad'' local minima.

\revision{For future developments, it would be interesting to understand whether infinite or long-range Hamiltonians can be used to boost QAOA performance on short-range models.
The idea is to add a further unitary $\nep^{-i\epsilon{\Ho}_z'}$ in Eq.~\eqref{eq:QAOA1}, generated by a long-range Hamiltonian $\Ho_z'$ unrelated to the problem to solve. 
This enlarges the portion of Hilbert space approximated with a QAOA Ansatz, at a fixed number of Trotter steps $\Ptrot$, at the price however of increasing the number of variational parameters. 
} 

At variance with Refs.~\cite{Zhou_PRX2020,Mbeng_arx19,Pagano_arXiv2019}, we are unable to construct minima in the energy landscape associated with smooth parameters $(\bgamma^*,\bbeta^*)$. 
\revision{These regular parameter choices are often desired since they are linked to adiabatic schedules, that can be used to infer the optimal protocol $s(t)$ in a continuous annealing scenario, and allow for a faster minimum search in the $2\Ptrot$-dimensional parameter space, once a solution for $\Ptrot' < \Ptrot$ is known~\cite{Pagano_arXiv2019,Mbeng_arx19}.}
Preliminary results \cite{Wauters_RL_arx2020} with reinforcement learning \cite{Sutton_2ed} methods applied to the QAOA evolution suggest however that smooth choices of 
$(\bgamma^*,\bbeta^*)$ do indeed exist, but they are hard to find with local optimizations. 
Whether {\em global} minima are related to smooth values of $(\bgamma^*,\bbeta^*)$ remains an open and interesting question.

\section*{Acknowledgments}
We acknowledge fruitful discussions with E. Panizon.
Research was partly supported by EU Horizon 2020 under ERC-ULTRADISS, Grant Agreement No. 834402.
GES acknowledges that his research has been conducted within the framework of the Trieste Institute for Theoretical Quantum Technologies (TQT).

\appendix

\newpage
\begin{widetext}
\section{Exact ground state preparation for $\Ptrot=1$}\label{app:P1}
In this appendix we will show that one can get the exact target ground state of the $\prm$-spin model with a single
QAOA step, $\Ptrot=1$, starting from the fully x-polarized state 
$|+\rangle= \frac{1}{\sqrt{2^{\Nsites}}} \bigotimes_{j=1}^{\Nsites} \left( \ket{\!\uparrow}_{j} + \ket{\!\downarrow}_j  \right)$,
provided the number of sites $\Nsites$ is {\em odd}.
This holds true for all possible values of $\prm$, and generalizes the result of Ref.~\onlinecite{Ho_arXiv2018} to $\prm>2$. 

\subsection{$\Ptrot=1$: requirements on $\beta$}
For $\Ptrot=1$ the QAOA state has only two parameters, which we will denote by $\gamma$ and $\beta$, without an index. 
Let $|\psitarget\rangle$ denote the (target) ground state of the model, and define the fidelity:
\begin{equation}\label{eqn:fidelity}
\calF(\gamma,\beta) = \Big| \braket{\psitarget}{\psi_{\Ptrot=1}(\gamma,\beta )}\Big|^2 = \left| \bra{\psitarget} \nep^{-i\beta \Ho_x} \nep^{-i \gamma \Ho_z}\ket{+}\right|^2 = 
\Big| \frac{1}{\sqrt{2^\Nsites}} \sum_l \nep^{-i\gamma E_l} \bra{\psitarget} \nep^{-i\beta \Ho_x} \ket{l} \Big|^2 \;,
\end{equation}
where we have expanded the initial state $|+\rangle =\frac{1}{\sqrt{2^\Nsites}} \sum_l \ket{l}$ as an equal superposition of all possible $2^{\Nsites}$ classical 
$\z$-basis configurations $\ket{l}$, and we used that $\Ho_z \ket{l} = E_l \ket{l}$, where $E_l$ is the energy of the configuration $\ket{l}$. 
Let us now define the following two $2^{\Nsites}$ dimensional complex vectors:
\begin{equation}
      (\vvect(\gamma))_l = \frac{1}{\sqrt{2^{\Nsites}}}\nep^{i \gamma E_l} \hspace{5mm} \mbox{and} \hspace{5mm} 
      (\uvect(\beta))_l      = \langle \psitarget | \nep^{-i\beta \Ho_x} \ket{l}  \;.
\end{equation}
Simple algebra shows that they have unit norm, $|| \vvect(\gamma) || = 1$ and $|| \uvect(\beta) || = 1$, and that  
the fidelity can be expressed as a scalar product of them:
$\calF(\gamma,\beta) =  | \vvect^{\dagger}(\gamma) \cdot \uvect(\beta) |^2$.
Hence, by the Cauchy-Schwarz inequality:
\begin{equation}
1 = \calF(\gamma,\beta)= |\vvect^\dagger(\gamma) \cdot \uvect(\beta) |^2 \iff \exists \, \theta \in \mathbb{R} \hspace{3mm} \mbox{such that} \hspace{3mm} \uvect(\beta)=\nep^{i\theta} \vvect(\gamma) \;,
\end{equation} 
i.e., the two vectors coincide, up to an overall phase factor. Since $| (\vvect(\gamma))_l |^2 = \frac{1}{2^{\Nsites}}$, this in turn implies that we must have
\begin{equation} \label{eqn:beta_requirement}
\Big| \langle \psitarget | \nep^{-i\beta \Ho_x} \ket{l} \Big|^2 = |(\uvect(\beta))_l|^2 = \frac{1}{2^{\Nsites}} \hspace{5mm} \forall l \;.
\end{equation}

So far, our arguments have been rather general. We now specialize our discussion to the case where $|\psitarget\rangle$ is the ground state of the classical $\prm$-spin ferromagnet.

For $\prm$ {\em odd}, we have $|\psitarget\rangle = | \! \uparrow \cdots \uparrow \rangle$, and a simple calculation shows that:
\begin{equation} \label{eqn:podd}
\langle \psitarget | \nep^{-i\beta \Ho_x} \ket{l} = \prod_{j=1}^{\Nsites} \langle \uparrow \! | \cos \beta \, \identity_j + i \sin \beta \, \PauliSigma^x_j | l_j \rangle 
= \big(\cos \beta\big)^{\Nup_l} \big(i \sin \beta\big)^{\Ndown_l} \;,
\end{equation}
where $\Nup_l$ and $\Ndown_l$ denote the number of $\uparrow$ and $\downarrow$ spins in the configuration $\ket{l}$. 
Hence the requirement given by Eq.~\eqref{eqn:beta_requirement} is satisfied only if 
\begin{equation}
\cos^2 \! \beta = \sin^2 \! \beta = \frac{1}{2} \hspace{5mm} \Longrightarrow \beta = \frac{\pi}{4}, \frac{3\pi}{4}, \frac{5\pi}{4}, \frac{7\pi}{4} \;.
\end{equation}
Similar arguments have been used, see Ref.~\onlinecite{Streif_arXiv19}, for the more general case in which $|\psitarget\rangle$ is the classical ground state of a generic spin-glass Hamiltonian.  

For $\prm$ {\em even} the calculation is slightly more involved, since the target state is now a non-classical superposition of the two degenerate ferromagnetic states
\begin{equation} \label{eqn:psitarget_peven}
|\psitarget\rangle = \frac{1}{\sqrt{2}} \big( |\! \uparrow \cdots \uparrow \rangle + | \! \downarrow \cdots \downarrow \rangle \big) \;.
\end{equation}
Hence:
\begin{equation}
\Big| \langle \psitarget | \nep^{-i\beta \Ho_x} \ket{l} \Big|^2 = \frac{1}{2} \Big|   \big(\cos \! \beta\big)^{\Nup_l}  \big(i \sin \! \beta\big)^{\Ndown_l} +  
\big(\cos \! \beta\big)^{\Ndown_l}  \big(i \sin \! \beta\big)^{\Nup_l} \Big|^2 \;. 
\end{equation}
Once again, one easily verifies that $\beta=\pifour$ satisfies the requirement \eqref{eqn:beta_requirement}, {\em provided $\Nsites$ is odd}, so that $\Nup_l$ and $\Ndown_l$ have
opposite parity and therefore $|i^{\Nup_l}+i^{\Ndown_l}|^2=2$. 

From now on we will therefore restrict our choice of $\beta$ to $\beta=\pifour$, a necessary condition for unit fidelity, and study the conditions that $\gamma$ has to verify.
Essentially, the value of $\gamma$ will have to be chosen in such a way that the various phase factors interfere constructively in a way that is independent of $l$.
To this goal, we notice that the energy $E_l$ of the configuration $\ket{l}$ can be expressed as:
\begin{equation}
E_l = - \langle l |\big( \sum_j \PauliSigma^z_j \big)^{\prm} | l \rangle = -(\Nup_l-\Ndown_l)^{\prm} = -\Magn_l^{\prm} \;,
\end{equation} 
where $\Magn_l = \Nup_l-\Ndown_l$ is the total magnetization of the configuration. 

\subsection{$\Ptrot=1$ and $\prm$ odd: requirements on $\gamma$}
Here we prove that for odd values of $\prm$ and $\Nsites$, the $\prm$-spin QAOA circuit of depth $\Ptrot=1$ and parameters $(\gamma = \pifour, \beta=\pifour)$ 
is sufficient to prepare the ferromagnetic target state $\ket{\psitarget}=|\uparrow \cdots \uparrow \rangle$. 
Substituting $\beta=\pifour$ and  $i=\nep^{i\frac{\pi}{2}}$ in Eq.~\eqref{eqn:podd}, and using $E_l=-\Magn_l^{\prm}$ in Eq.~\eqref{eqn:fidelity} we get:
\begin{eqnarray}\label{eq:fidPifour}
\calF(\gamma,\pifour) &=& \Big| \frac{1}{2^\Nsites} \sum_l \nep^{i \gamma \Magn_l^\prm} \nep^{i \frac{\pi}{2}\Ndown_l} \Big|^2 \;.
\end{eqnarray}
Taking $\gamma=\pifour$ gives
\begin{eqnarray}
\calF(\pifour,\pifour)
&=& \left|\frac{1}{2^\Nsites}\sum_l \exp \left(i \frac{2\pi}{8} \left[\left(\Magn_l^\prm + 2\Ndown_l \right) \mod 8 \right] \right)\right|^2 \nonumber\\
&=& \left|\frac{1}{2^\Nsites}\sum_l \exp \left(i \frac{2\pi}{8} \left[\left(\Magn_l + 2\Ndown_l \right) \mod 8 \right] \right)\right|^2 \nonumber\\
&=& \left|\frac{1}{2^\Nsites}\sum_l \exp \left(i \frac{2\pi}{8} \left[\Nsites \mod 8 \right]\right)\right|^2 = 1 \;,
\label{eqn:fidelity_p_odd_(pi/4,pi/4)}
\end{eqnarray}
where we have used the fact that for $\Nsites$ odd, $\Magn_l=\Nsites-2\Ndown_l$ is also odd and the following property of arithmetic congruences holds:
\begin{equation}
\Magn_l^{\prm-1} = 1 \mod 8 \hspace{3mm} \Longrightarrow \hspace{3mm} \Magn_l^{\prm} = \Magn_l \mod 8  \hspace{10mm} \mbox{if } \prm \mbox{ is odd}\;.
\end{equation}

%

Eq.~\eqref{eqn:fidelity_p_odd_(pi/4,pi/4)} proves our initial claim, that the QAOA protocol $(\gamma = \pifour, \beta=\pifour)$ 
prepares the target ground state of $\Ho_z$ for the $\prm$-spin ferromagnet with unit fidelity, provided $\Nsites$ and $\prm$ are both odd.

\subsection{$\Ptrot=1$ and $\prm$ even: requirements on $\gamma$}
For even values of $\prm$, the system is $\Ztwo$ symmetric. The $\prm$-spin  QAOA circuit preserves such symmetry. 
Therefore, the targeted ground state of $\Ho_z$ is $\ket{\psitarget}$ in Eq.~\eqref{eqn:psitarget_peven}.  
As we did in the previous section, we compute the fidelity between the output $\ket{\psi_{\Ptrot=1}(\gamma,\beta=\pifour )}$ of the QAOA circuit and the (non-classical) target state 
$\ket{\psitarget}$ in Eq.~\eqref{eqn:psitarget_peven}:
\begin{equation} \label{eqn:fidelity_p_even_with_beta}
\calF(\gamma,\beta=\pifour) = \left| \frac{1}{2^{\Nsites}} \sum_l \nep^{i \gamma \Magn_l^{\prm}} 
\left( \frac{\nep^{i\frac{\pi}{2} \Ndown_l} + \nep^{i\frac{\pi}{2} \Nup_l}}{\sqrt{2}} \right)  \right|^2 \;.
\end{equation}
We observe that for $\Nsites=\Nup_l+\Ndown_l$ odd, $\Nup_l$ and $ \Ndown_l$ must have {\em opposite parity}, and the term inside parenthesis is a
pure phase factor, which can be expressed as:
\begin{equation} \label{eqn:phase}
\left( \frac{\nep^{i\frac{\pi}{2} \Ndown_l} + \nep^{i\frac{\pi}{2} \Nup_l}}{\sqrt{2}} \right)  = \nep^{i\frac{\pi}{4} \Nsites} \nep^{-i \pi f(\Magn_l)} \;, 
\end{equation}
where 
\begin{eqnarray} \label{eqn:f_def}
f(\Magn) = \begin{cases}
0 	& \mbox{for }\;\; \Magn \mod 8 = \pm 1 \vspace{3mm} \\
1 	& \mbox{for }\;\; \Magn \mod 8 = \pm 3  
\end{cases}\;.
\end{eqnarray}
Hence, omitting the irrelevant $l$-independent common factor $\nep^{i\frac{\pi}{4} \Nsites}$, we can rewrite the fidelity as:
\begin{equation} \label{eqn:fidelity_p_even_with_f}
\calF(\gamma,\pifour) = \Big| \frac{1}{2^{\Nsites}} \sum_l \nep^{i \left( \gamma \Magn_l^{\prm} - \pi f(\Magn_l) \right)} \Big|^2 \;.
\end{equation}
The arithmetics to prove that the various phase factors can be made $l$-independent for a judicious choice of $\gamma$ is now, for even $\prm$, slightly more
involved. By experimenting with this expression for  $\prm\leq 10$, we have come out with the following unconventional parameterization of an even value of $\prm$:
for every even $\prm$, two natural numbers $n$ and $k$ can be found such that:
\begin{equation} \label{eqn:p_param}
\prm = 2^{k+1} + n 2^k \;.
\end{equation}
Correspondingly, given the value of $k$ in Eq.~\eqref{eqn:p_param}, we will set the value of $\gamma$ to:
\begin{equation} 
\gamma_k = \frac{2\pi}{2^{k+4}} \;.
\end{equation}
The crucial arithmetic identity which we will use --- see Sec.~\ref{sec:proof} for a proof --- is the following:
\begin{equation} \label{eqn:identity}
m^{2^{k+1}+n2^{k}} \mod 2^{k+4} = f(m) \, 2^{k+3} + 1 \hspace{15mm} \forall  \mbox{    $m\in \mathbb{Z}$    with     $m$ odd} \;,
\end{equation}
where $f(m)$ is the function given in Eq.~\eqref{eqn:f_def}.

With these definitions, it is immediate to verify that: 
\begin{eqnarray} \label{eqn:fidelity_peven_parameterized}
\calF(\gamma_k,\pifour) 
&=&  \left| \frac{1}{2^{\Nsites}} \sum_l \nep^{-i \pi f(\Magn_l)} \exp\left(i \frac{2\pi}{2^{k+4}} \left(\Magn_l\right)^{2^{k+1}+n2^{k}}\right) \right|^2\nonumber\\
&=&  \left| \frac{1}{2^{\Nsites}} \sum_l \nep^{-i \pi f(\Magn_l)} \exp\left(i \frac{2\pi}{2^{k+4}} \left[ \left(\Magn_l\right)^{2^{k+1}+n2^{k}} \mod 2^{k+4} \right] \right) \right|^2\nonumber\\
&=& \left| \frac{1}{2^{\Nsites}} \sum_l \nep^{-i \pi f(\Magn_l)} \exp\left(i \frac{2\pi}{2^{k+4}} \left( f(\Magn_l) \, 2^{k+3}  +1 \right) \right) \right|^2\nonumber\\
&=& \left| \frac{1}{2^{\Nsites}} \sum_l \nep^{-i \pi f(\Magn_l)} \nep^{i \pi f(\Magn_l) + i \frac{2\pi}{2^{k+4}} } \right|^2 = 1 \;. 
\end{eqnarray}

\paragraph{Proof of identity in Eq.~\eqref{eqn:identity}} \label{sec:proof}
For completeness, we also present a proof of the arithmetic identity Eq.~\eqref{eqn:identity}. 
To prove Eq.~\eqref{eqn:identity}, it is sufficient to show that
\begin{eqnarray}\label{eqn:mod_2^k_eq_biss}
\forall k\in \mathbb{N}, m\in \mathbb{Z}, m \mbox{ odd}: \;\;\;\;  (m^{2^{k+1}} - 1)\mod 2^{k+4} &=& f(m) \, 2^{k+3}\;.
\end{eqnarray}
We prove Eq.~\eqref{eqn:mod_2^k_eq_biss} by induction over $k$:
\begin{itemize}
    \item[(i)] We show that Eq.~\eqref{eqn:mod_2^k_eq_biss} holds for $k=0$.
    
    For $k = 0$, a direct computation,  for odd $m$,  gives:
    \begin{eqnarray}
    (m^{2^{0+1}}-1) \mod 2^{0+4}&=& (m^2 - 1) \mod 16\nonumber\\
    &=&  (m- 1)(m+1) \mod 16\nonumber\\
    &=& f(m) \, 2^3
    \end{eqnarray} 
    
    \item[(ii)] We show that if Eq.~\eqref{eqn:mod_2^k_eq_biss} holds for a given $k\in \mathbb{N}$ and for all odd $m\in\mathbb{N}$, then it holds also for $k+1$. 
    
    Using Eq.~\eqref{eqn:mod_2^k_eq_biss}, we write
    \begin{eqnarray}
    m^{2^{k+1}} &=& a_m 2^{k+4} + f(m)\, 2^{k+3} + 1 \;,
    \end{eqnarray}
    with $a_m\in \mathbb{Z}$. Then, we have
    \begin{eqnarray}
    (m^{2^{(k+1)+1}}-1)&=& (m^{2^{k+1}}-1) (m^{2^{k+1}}+1)\nonumber\\
    &=&(a_m 2^{k+4} + f(m)\, 2^{k+3})(a_m 2^{k+4} + f(m)\, 2^{k+3} + 2)\nonumber\\
    &=&(a_m 2^{k+5} + f(m)\, 2^{k+4})(a_m 2^{k+3} + f(m)\, 2^{k+2} + 1) \;.
    \end{eqnarray}
    From this, we derive
    \begin{eqnarray}
    (m^{2^{(k+1)+1}}-1) \mod 2^{(k+1)+4} 
    &=& f(m)\, 2^{k+4} (a_m 2^{k+3} + f(m)\, 2^{k+2} + 1)  \mod 2^{k+5} \\
    &=& f(m)\, 2^{k+4}
    \;,
    \end{eqnarray}
    where we have used that $f(m)=0,1$ for all odd $m\in\mathbb{Z}$. This indeed implies that for all $k\in \mathbb{N}$:
    \begin{eqnarray}
    (m^{2^{k+1}}-1) \mod 2^{k+4} = f(m)\, 2^{k+3} \implies (m^{2^{(k+1)+1}}-1) \mod 2^{(k+1) +4} = f(m)\, 2^{(k+1)+3}\;.
    \end{eqnarray}
    This concludes the proof by induction of Eq.~\eqref{eqn:mod_2^k_eq_biss}.
\end{itemize}

Incidentally, as an immediate consequence of Eq.~\eqref{eqn:mod_2^k_eq_biss} we get that, for any $n\in \mathbb{N}$:
        \begin{eqnarray}
        m^{2^{k+1}} \mod 2^{k+4} &=& f(m) \, 2^{k{+3}} + 1\\
        m^{2^{k{+2}}}\mod 2^{k{+4}} &=& 1 \label{eqn:mod_2^k_Gauss}\\
        m^{2^{k{+1}}+n2^{k}} \mod 2^{k{+4}} &=& f(m) \, 2^{k{+3}} + 1\;.
        \end{eqnarray} 
Notice that Eq.~\eqref{eqn:mod_2^k_Gauss} also follows from the properties of the multiplicative group of integers modulo $2^k$ discussed in Refs.~\onlinecite{book:Gauss1966,wiki:mult_groups} (eg.  $(\mathbb{Z}/2^{k{+4}}\mathbb{Z})^{\times}\cong C_2 \times C_{2^{k{+2}}}$).


\section{Symmetries of the parameter space for general $\Ptrot$, $\Nsites$ and $\prm$} \label{app:sym}
We discuss here the symmetries in the parameter space of Eq.~\eqref{eq:QAOA_energy}, which we recall here for convenience
\begin{equation}
E_{\Ptrot}(\bgamma,\bbeta )=\langle \psi_{\Ptrot}(\bgamma,\bbeta)| \Htarg |\psi_{\Ptrot}(\bgamma,\bbeta) \rangle \;.
\end{equation}
A first trivial operation that leaves the energy unaltered is the inversion $(\bgamma,\bbeta) \to -(\bgamma,\bbeta)$, which corresponds to the complex conjugate of Eq.~\eqref{eq:QAOA_energy}.
Indeed it is immediate to see that 
\begin{equation}
|\psi_{\Ptrot}(-\bgamma,-\bbeta )\rangle 
= \prod_{m=1}^{\Ptrot} \nep^{i \beta_m \Hx}\nep^{i\gamma_m \Hz} |\psi_0\rangle = |\psi_{\Ptrot}(\bgamma,\bbeta )\rangle^*\;,
\end{equation}
given that $\ket{\psi_0}=\ket{+}$ is a real wavefunction in the basis of $\Spin_z$.
 
The symmetries on the $\bbeta$ parameters are shared by all QAOA wavefunctions where quantum fluctuations are induced by a magnetic field transverse 
to the computational basis. 
We can write a single evolution operator $\nep^{-i \beta_m \Hx}$ as a set of rotation on each individual spin
\begin{equation}\label{eqn:Sx}
\nep^{-i \beta_m \Hx}= \nep^{i \beta_m \sum_{j=1}^\Nsites \PauliSigma^x_j} = \bigotimes_{j=1}^\Nsites \left( \cos \beta_m + \PauliSigma^x_j \sin \beta_m \right) \;.
\end{equation}
A shift $\beta_m \to \beta_m + \pi$ changes the sign of each term in the product, leading to 
\begin{equation}
\nep^{-i (\beta_m+\pi) \Hx} =  \bigotimes_{j=1}^\Nsites \left( -\cos \beta_m  - \PauliSigma^x_j \sin \beta_m  \right) = 
(-1)^\Nsites \bigotimes_{j=1}^\Nsites \left( \cos \beta_m + \PauliSigma^x_j \sin \beta_m \right) \;,
\end{equation}
which is a trivial global phase that does not change the energy in Eq.~\eqref{eq:QAOA_energy}.
Moreover, if $\prm$ is even, the target Hamiltonian is $\Ztwo$ symmetric. 
Recall that $ \Hx=-\Spin_x$ (twice the total spin), which implies that: 
\begin{equation}
\nep^{i\frac{\pi}{2} \Spin_x} \Htarg \nep^{-i\frac{\pi}{2} \Spin_x} = \Htarg \;,
\end{equation}
because $\nep^{-i\frac{\pi}{2} \Spin_x}$ is a $\pi$-rotation around the $\x$-direction which gives a global spin flip $\PauliSigma^z_j \to - \PauliSigma^z_j$, 
leading to $E_{\Ptrot}(\bgamma,\bbeta+\frac{\pi}{2}) = E_{\Ptrot}(\bgamma,\bbeta)$.

The symmetry for $\bgamma$ is subtler and is model-specific.
Notice first that $\Spin_z=\sum_j \PauliSigma^z_j$ has integer eigenvalues, even or odd depending on $\Nsites$, and so does $\Hz=-\Spin_z^{\prm}$.
Following the same notation introduced previously, we write a single QAOA evolution operator as
\begin{equation}
\nep^{ -i \gamma_m \Ho_z } = \sum_l \nep^{i \gamma_m \Magn^\prm_l } \ket{l} \bra{l} \;.
\end{equation}
If $\Nsites$ is odd the eigenvalues $\Magn^\prm_l$ of $\Spin_z^\prm$ are also odd, and the periodicity of $\gamma_m$ is $\pi$, because
\begin{equation}
(\gamma_m + \pi) \Magn^\prm_l = \gamma_m \Magn^\prm_l + \pi \mod 2\pi \ .
\end{equation}
Hence the shift $\gamma_m \to \gamma_m+\pi$ introduces a global phase $\nep^{ -i (\gamma_m+\pi) \Ho_z } = - \nep^{ -i \gamma_m \Ho_z }$, 
which is irrelevant in the expectation value of the energy.
If $\Nsites$ is even, the eigenvalues $\Magn_l^\prm$ of $\Spin_z^{\prm}$ are multiples of $2^\prm$, hence
\begin{equation}
(\gamma_m + \frac{\pi}{2^{\prm-1}}) \Magn^\prm_l = \gamma_m \Magn^\prm_l \mod 2\pi \;,
\end{equation}
which means that $\nep^{i(\gamma_m+\frac{\pi}{2^{\prm-1}})\Spin_z^\prm} = \nep^{i \gamma_m\Spin_z^\prm}$.
In table \ref{tab:sym} we summarize the symmetries we have discussed. 

\begin{table}
\begin{center}
\begin{tabular}{ l | l } 
 $\forall \ \Nsites, \ \prm$ & \hspace{3mm} $E(-\bgamma,-\bbeta)=E(\bgamma,\bbeta)$ \\
 \hline
 $\prm$ odd & \hspace{3mm} $E(\bgamma,\bbeta+\pi)=E(\bgamma,\bbeta)$ \\
 \hline
 $\prm$ even & \hspace{3mm} $E(\bgamma,\bbeta+\frac{\pi}{2})=E(\bgamma,\bbeta)$ \\
 \hline
 $\Nsites$ odd & \hspace{3mm} $E(\bgamma+\pi,\bbeta)=E(\bgamma,\bbeta)$ \\
 \hline
 $\Nsites$ even & \hspace{3mm} $E(\bgamma + \frac{\pi}{2^{\prm-1}},\bbeta)=E(\bgamma,\bbeta))$
\end{tabular}
\caption{Symmetry operations for the QAOA process of the p-spin model. It is understood that any component of $\bgamma$ or $\bbeta$ can be modified.}\label{tab:sym}
\end{center}
\end{table}
\end{widetext}


\begin{thebibliography}{54}
\expandafter\ifx\csname natexlab\endcsname\relax\def\natexlab#1{#1}\fi
\expandafter\ifx\csname bibnamefont\endcsname\relax
  \def\bibnamefont#1{#1}\fi
\expandafter\ifx\csname bibfnamefont\endcsname\relax
  \def\bibfnamefont#1{#1}\fi
\expandafter\ifx\csname citenamefont\endcsname\relax
  \def\citenamefont#1{#1}\fi
\expandafter\ifx\csname url\endcsname\relax
  \def\url#1{\texttt{#1}}\fi
\expandafter\ifx\csname urlprefix\endcsname\relax\def\urlprefix{URL }\fi
\providecommand{\bibinfo}[2]{#2}
\providecommand{\eprint}[2][]{\url{#2}}

\bibitem[{\citenamefont{Ac{\'{\i}}n et~al.}(2018)\citenamefont{Ac{\'{\i}}n,
  Bloch, Buhrman, Calarco, Eichler, Eisert, Esteve, Gisin, Glaser, Jelezko
  et~al.}}]{QuantumTech_2018}
\bibinfo{author}{\bibfnamefont{A.}~\bibnamefont{Ac{\'{\i}}n}},
  \bibinfo{author}{\bibfnamefont{I.}~\bibnamefont{Bloch}},
  \bibinfo{author}{\bibfnamefont{H.}~\bibnamefont{Buhrman}},
  \bibinfo{author}{\bibfnamefont{T.}~\bibnamefont{Calarco}},
  \bibinfo{author}{\bibfnamefont{C.}~\bibnamefont{Eichler}},
  \bibinfo{author}{\bibfnamefont{J.}~\bibnamefont{Eisert}},
  \bibinfo{author}{\bibfnamefont{D.}~\bibnamefont{Esteve}},
  \bibinfo{author}{\bibfnamefont{N.}~\bibnamefont{Gisin}},
  \bibinfo{author}{\bibfnamefont{S.~J.} \bibnamefont{Glaser}},
  \bibinfo{author}{\bibfnamefont{F.}~\bibnamefont{Jelezko}},
  \bibnamefont{et~al.}, \bibinfo{journal}{New Journal of Physics}
  \textbf{\bibinfo{volume}{20}}, \bibinfo{pages}{080201}
  (\bibinfo{year}{2018}),
  \urlprefix\url{https://doi.org/10.1088%2F1367-2630%2Faad1ea}.

\bibitem[{\citenamefont{Kurizki et~al.}(2015)\citenamefont{Kurizki, Bertet,
  Kubo, M{\o}lmer, Petrosyan, Rabl, and Schmiedmayer}}]{Kurizki_PNAS15}
\bibinfo{author}{\bibfnamefont{G.}~\bibnamefont{Kurizki}},
  \bibinfo{author}{\bibfnamefont{P.}~\bibnamefont{Bertet}},
  \bibinfo{author}{\bibfnamefont{Y.}~\bibnamefont{Kubo}},
  \bibinfo{author}{\bibfnamefont{K.}~\bibnamefont{M{\o}lmer}},
  \bibinfo{author}{\bibfnamefont{D.}~\bibnamefont{Petrosyan}},
  \bibinfo{author}{\bibfnamefont{P.}~\bibnamefont{Rabl}}, \bibnamefont{and}
  \bibinfo{author}{\bibfnamefont{J.}~\bibnamefont{Schmiedmayer}},
  \bibinfo{journal}{Proc. Natl. Acad. Sci. USA} \textbf{\bibinfo{volume}{112}},
  \bibinfo{pages}{3866} (\bibinfo{year}{2015}), ISSN \bibinfo{issn}{0027-8424}.

\bibitem[{\citenamefont{Preskill}(2012)}]{Preskill_arXiv2012}
\bibinfo{author}{\bibfnamefont{J.}~\bibnamefont{Preskill}},
  \bibinfo{journal}{arXiv e-prints} p. \bibinfo{pages}{arXiv:1203.5813}
  (\bibinfo{year}{2012}), \eprint{1203.5813}.

\bibitem[{\citenamefont{{Farhi} and {Harrow}}(2016)}]{Farhi_arXiv2016}
\bibinfo{author}{\bibfnamefont{E.}~\bibnamefont{{Farhi}}} \bibnamefont{and}
  \bibinfo{author}{\bibfnamefont{A.~W.} \bibnamefont{{Harrow}}},
  \bibinfo{journal}{arXiv e-prints} \bibinfo{eid}{arXiv:1602.07674}
  (\bibinfo{year}{2016}), \eprint{1602.07674}.

\bibitem[{\citenamefont{Arute et~al.}(2019)\citenamefont{Arute, Arya, Babbush,
  Bacon, Bardin, Barends, Biswas, Boixo, Brandao, Buell
  et~al.}}]{Nature_QuantumSupremacy}
\bibinfo{author}{\bibfnamefont{F.}~\bibnamefont{Arute}},
  \bibinfo{author}{\bibfnamefont{K.}~\bibnamefont{Arya}},
  \bibinfo{author}{\bibfnamefont{R.}~\bibnamefont{Babbush}},
  \bibinfo{author}{\bibfnamefont{D.}~\bibnamefont{Bacon}},
  \bibinfo{author}{\bibfnamefont{J.~C.} \bibnamefont{Bardin}},
  \bibinfo{author}{\bibfnamefont{R.}~\bibnamefont{Barends}},
  \bibinfo{author}{\bibfnamefont{R.}~\bibnamefont{Biswas}},
  \bibinfo{author}{\bibfnamefont{S.}~\bibnamefont{Boixo}},
  \bibinfo{author}{\bibfnamefont{F.~G. S.~L.} \bibnamefont{Brandao}},
  \bibinfo{author}{\bibfnamefont{D.~A.} \bibnamefont{Buell}},
  \bibnamefont{et~al.}, \bibinfo{journal}{Nature}
  \textbf{\bibinfo{volume}{574}}, \bibinfo{pages}{505} (\bibinfo{year}{2019}).

\bibitem[{\citenamefont{Lloyd}(1996)}]{Lloyd_SCI96}
\bibinfo{author}{\bibfnamefont{S.}~\bibnamefont{Lloyd}},
  \bibinfo{journal}{Science} \textbf{\bibinfo{volume}{273}},
  \bibinfo{pages}{1073} (\bibinfo{year}{1996}), ISSN \bibinfo{issn}{00368075,
  10959203}.

\bibitem[{\citenamefont{Buluta and Nori}(2009)}]{Buluta_SCI09}
\bibinfo{author}{\bibfnamefont{I.}~\bibnamefont{Buluta}} \bibnamefont{and}
  \bibinfo{author}{\bibfnamefont{F.}~\bibnamefont{Nori}},
  \bibinfo{journal}{Science} \textbf{\bibinfo{volume}{326}},
  \bibinfo{pages}{108} (\bibinfo{year}{2009}), ISSN \bibinfo{issn}{0036-8075}.

\bibitem[{\citenamefont{Georgescu et~al.}(2014)\citenamefont{Georgescu, Ashhab,
  and Nori}}]{Nori_RMP14}
\bibinfo{author}{\bibfnamefont{I.~M.} \bibnamefont{Georgescu}},
  \bibinfo{author}{\bibfnamefont{S.}~\bibnamefont{Ashhab}}, \bibnamefont{and}
  \bibinfo{author}{\bibfnamefont{F.}~\bibnamefont{Nori}},
  \bibinfo{journal}{Rev. Mod. Phys.} \textbf{\bibinfo{volume}{86}},
  \bibinfo{pages}{153} (\bibinfo{year}{2014}).

\bibitem[{\citenamefont{Lucas}(2014)}]{Lucas_FrontPhys2014}
\bibinfo{author}{\bibfnamefont{A.}~\bibnamefont{Lucas}},
  \bibinfo{journal}{Frontiers in Physics} \textbf{\bibinfo{volume}{2}},
  \bibinfo{pages}{5} (\bibinfo{year}{2014}).

\bibitem[{\citenamefont{Finnila et~al.}(1994)\citenamefont{Finnila, Gomez,
  Sebenik, Stenson, and Doll}}]{Finnila_CPL94}
\bibinfo{author}{\bibfnamefont{A.~B.} \bibnamefont{Finnila}},
  \bibinfo{author}{\bibfnamefont{M.~A.} \bibnamefont{Gomez}},
  \bibinfo{author}{\bibfnamefont{C.}~\bibnamefont{Sebenik}},
  \bibinfo{author}{\bibfnamefont{C.}~\bibnamefont{Stenson}}, \bibnamefont{and}
  \bibinfo{author}{\bibfnamefont{J.~D.} \bibnamefont{Doll}},
  \bibinfo{journal}{Chem. Phys. Lett.} \textbf{\bibinfo{volume}{219}},
  \bibinfo{pages}{343} (\bibinfo{year}{1994}).

\bibitem[{\citenamefont{Kadowaki and Nishimori}(1998)}]{Kadowaki_PRE98}
\bibinfo{author}{\bibfnamefont{T.}~\bibnamefont{Kadowaki}} \bibnamefont{and}
  \bibinfo{author}{\bibfnamefont{H.}~\bibnamefont{Nishimori}},
  \bibinfo{journal}{Phys. Rev. E} \textbf{\bibinfo{volume}{58}},
  \bibinfo{pages}{5355} (\bibinfo{year}{1998}).

\bibitem[{\citenamefont{Brooke et~al.}(1999)\citenamefont{Brooke, Bitko,
  Rosenbaum, and Aeppli}}]{Brooke_SCI99}
\bibinfo{author}{\bibfnamefont{J.}~\bibnamefont{Brooke}},
  \bibinfo{author}{\bibfnamefont{D.}~\bibnamefont{Bitko}},
  \bibinfo{author}{\bibfnamefont{T.~F.} \bibnamefont{Rosenbaum}},
  \bibnamefont{and} \bibinfo{author}{\bibfnamefont{G.}~\bibnamefont{Aeppli}},
  \bibinfo{journal}{Science} \textbf{\bibinfo{volume}{284}},
  \bibinfo{pages}{779} (\bibinfo{year}{1999}).

\bibitem[{\citenamefont{Santoro et~al.}(2002)\citenamefont{Santoro,
  {Marto\v{n}\'{a}k}, Tosatti, and Car}}]{Santoro_SCI02}
\bibinfo{author}{\bibfnamefont{G.~E.} \bibnamefont{Santoro}},
  \bibinfo{author}{\bibfnamefont{R.}~\bibnamefont{{Marto\v{n}\'{a}k}}},
  \bibinfo{author}{\bibfnamefont{E.}~\bibnamefont{Tosatti}}, \bibnamefont{and}
  \bibinfo{author}{\bibfnamefont{R.}~\bibnamefont{Car}},
  \bibinfo{journal}{Science} \textbf{\bibinfo{volume}{295}},
  \bibinfo{pages}{2427} (\bibinfo{year}{2002}).

\bibitem[{\citenamefont{Farhi et~al.}(2001)\citenamefont{Farhi, Goldstone,
  Gutmann, Lapan, Lundgren, and Preda}}]{Farhi_SCI01}
\bibinfo{author}{\bibfnamefont{E.}~\bibnamefont{Farhi}},
  \bibinfo{author}{\bibfnamefont{J.}~\bibnamefont{Goldstone}},
  \bibinfo{author}{\bibfnamefont{S.}~\bibnamefont{Gutmann}},
  \bibinfo{author}{\bibfnamefont{J.}~\bibnamefont{Lapan}},
  \bibinfo{author}{\bibfnamefont{A.}~\bibnamefont{Lundgren}}, \bibnamefont{and}
  \bibinfo{author}{\bibfnamefont{D.}~\bibnamefont{Preda}},
  \bibinfo{journal}{Science} \textbf{\bibinfo{volume}{292}},
  \bibinfo{pages}{472} (\bibinfo{year}{2001}).

\bibitem[{\citenamefont{Albash and Lidar}(2018{\natexlab{a}})}]{Albash_RMP18}
\bibinfo{author}{\bibfnamefont{T.}~\bibnamefont{Albash}} \bibnamefont{and}
  \bibinfo{author}{\bibfnamefont{D.~A.} \bibnamefont{Lidar}},
  \bibinfo{journal}{Rev. Mod. Phys.} \textbf{\bibinfo{volume}{90}},
  \bibinfo{pages}{015002} (\bibinfo{year}{2018}{\natexlab{a}}).

\bibitem[{\citenamefont{Heim et~al.}(2015)\citenamefont{Heim, R{\o}nnow,
  Isakov, and Troyer}}]{Heim_SCI15}
\bibinfo{author}{\bibfnamefont{B.}~\bibnamefont{Heim}},
  \bibinfo{author}{\bibfnamefont{T.~F.} \bibnamefont{R{\o}nnow}},
  \bibinfo{author}{\bibfnamefont{S.~V.} \bibnamefont{Isakov}},
  \bibnamefont{and} \bibinfo{author}{\bibfnamefont{M.}~\bibnamefont{Troyer}},
  \bibinfo{journal}{Science} \textbf{\bibinfo{volume}{348}},
  \bibinfo{pages}{215} (\bibinfo{year}{2015}).

\bibitem[{\citenamefont{{Crosson} and {Harrow}}(2016)}]{Crosson_FOCS2016}
\bibinfo{author}{\bibfnamefont{E.}~\bibnamefont{{Crosson}}} \bibnamefont{and}
  \bibinfo{author}{\bibfnamefont{A.~W.} \bibnamefont{{Harrow}}}, in
  \emph{\bibinfo{booktitle}{2016 IEEE 57th Annual Symposium on Foundations of
  Computer Science (FOCS)}} (\bibinfo{year}{2016}), pp.
  \bibinfo{pages}{714--723}, ISSN \bibinfo{issn}{0272-5428}.

\bibitem[{\citenamefont{Denchev et~al.}(2016)\citenamefont{Denchev, Boixo,
  Isakov, Ding, Babbush, Smelyanskiy, Martinis, and Neven}}]{Naven_PRX2016}
\bibinfo{author}{\bibfnamefont{V.~S.} \bibnamefont{Denchev}},
  \bibinfo{author}{\bibfnamefont{S.}~\bibnamefont{Boixo}},
  \bibinfo{author}{\bibfnamefont{S.~V.} \bibnamefont{Isakov}},
  \bibinfo{author}{\bibfnamefont{N.}~\bibnamefont{Ding}},
  \bibinfo{author}{\bibfnamefont{R.}~\bibnamefont{Babbush}},
  \bibinfo{author}{\bibfnamefont{V.}~\bibnamefont{Smelyanskiy}},
  \bibinfo{author}{\bibfnamefont{J.}~\bibnamefont{Martinis}}, \bibnamefont{and}
  \bibinfo{author}{\bibfnamefont{H.}~\bibnamefont{Neven}},
  \bibinfo{journal}{Phys. Rev. X} \textbf{\bibinfo{volume}{6}},
  \bibinfo{pages}{031015} (\bibinfo{year}{2016}).

\bibitem[{\citenamefont{Albash and Lidar}(2018{\natexlab{b}})}]{Albash_PRX2018}
\bibinfo{author}{\bibfnamefont{T.}~\bibnamefont{Albash}} \bibnamefont{and}
  \bibinfo{author}{\bibfnamefont{D.~A.} \bibnamefont{Lidar}},
  \bibinfo{journal}{Phys. Rev. X} \textbf{\bibinfo{volume}{8}},
  \bibinfo{pages}{031016} (\bibinfo{year}{2018}{\natexlab{b}}).

\bibitem[{\citenamefont{Jörg et~al.}(2010)\citenamefont{Jörg, Krzakala,
  Kurchan, Maggs, and Pujos}}]{Jorg_EPL10}
\bibinfo{author}{\bibfnamefont{T.}~\bibnamefont{Jörg}},
  \bibinfo{author}{\bibfnamefont{F.}~\bibnamefont{Krzakala}},
  \bibinfo{author}{\bibfnamefont{J.}~\bibnamefont{Kurchan}},
  \bibinfo{author}{\bibfnamefont{A.~C.} \bibnamefont{Maggs}}, \bibnamefont{and}
  \bibinfo{author}{\bibfnamefont{J.}~\bibnamefont{Pujos}},
  \bibinfo{journal}{EPL} \textbf{\bibinfo{volume}{89}}, \bibinfo{pages}{40004}
  (\bibinfo{year}{2010}).

\bibitem[{\citenamefont{Bapst and Semerjian}(2012)}]{Bapst_JSTAT12}
\bibinfo{author}{\bibfnamefont{V.}~\bibnamefont{Bapst}} \bibnamefont{and}
  \bibinfo{author}{\bibfnamefont{G.}~\bibnamefont{Semerjian}},
  \bibinfo{journal}{JSTAT} p. \bibinfo{pages}{P06007} (\bibinfo{year}{2012}).

\bibitem[{\citenamefont{Wauters et~al.}(2017)\citenamefont{Wauters, Fazio,
  Nishimori, and Santoro}}]{Wauters_PRA17}
\bibinfo{author}{\bibfnamefont{M.~M.} \bibnamefont{Wauters}},
  \bibinfo{author}{\bibfnamefont{R.}~\bibnamefont{Fazio}},
  \bibinfo{author}{\bibfnamefont{H.}~\bibnamefont{Nishimori}},
  \bibnamefont{and} \bibinfo{author}{\bibfnamefont{G.~E.}
  \bibnamefont{Santoro}}, \bibinfo{journal}{Phys. Rev. A}
  \textbf{\bibinfo{volume}{96}}, \bibinfo{pages}{022326}
  (\bibinfo{year}{2017}).

\bibitem[{\citenamefont{Nishimori and Takada}(2017)}]{Nishimori_ICT2017}
\bibinfo{author}{\bibfnamefont{H.}~\bibnamefont{Nishimori}} \bibnamefont{and}
  \bibinfo{author}{\bibfnamefont{K.}~\bibnamefont{Takada}},
  \bibinfo{journal}{Frontiers in ICT} \textbf{\bibinfo{volume}{4}},
  \bibinfo{pages}{2} (\bibinfo{year}{2017}), ISSN \bibinfo{issn}{2297-198X}.

\bibitem[{\citenamefont{Susa et~al.}(2018)\citenamefont{Susa, Yamashiro,
  Yamamoto, and Nishimori}}]{Nishimori_JPSJ18}
\bibinfo{author}{\bibfnamefont{Y.}~\bibnamefont{Susa}},
  \bibinfo{author}{\bibfnamefont{Y.}~\bibnamefont{Yamashiro}},
  \bibinfo{author}{\bibfnamefont{M.}~\bibnamefont{Yamamoto}}, \bibnamefont{and}
  \bibinfo{author}{\bibfnamefont{H.}~\bibnamefont{Nishimori}},
  \bibinfo{journal}{J. Phys. Soc. Jpn.} \textbf{\bibinfo{volume}{87}},
  \bibinfo{pages}{023002} (\bibinfo{year}{2018}).

\bibitem[{\citenamefont{Passarelli et~al.}(2019)\citenamefont{Passarelli,
  Cataudella, and Lucignano}}]{Passarelli_PRB19}
\bibinfo{author}{\bibfnamefont{G.}~\bibnamefont{Passarelli}},
  \bibinfo{author}{\bibfnamefont{V.}~\bibnamefont{Cataudella}},
  \bibnamefont{and}
  \bibinfo{author}{\bibfnamefont{P.}~\bibnamefont{Lucignano}},
  \bibinfo{journal}{Phys. Rev. B} \textbf{\bibinfo{volume}{100}},
  \bibinfo{pages}{024302} (\bibinfo{year}{2019}).

\bibitem[{\citenamefont{Passarelli et~al.}(2018)\citenamefont{Passarelli,
  De~Filippis, Cataudella, and Lucignano}}]{Passarelli_PRA19}
\bibinfo{author}{\bibfnamefont{G.}~\bibnamefont{Passarelli}},
  \bibinfo{author}{\bibfnamefont{G.}~\bibnamefont{De~Filippis}},
  \bibinfo{author}{\bibfnamefont{V.}~\bibnamefont{Cataudella}},
  \bibnamefont{and}
  \bibinfo{author}{\bibfnamefont{P.}~\bibnamefont{Lucignano}},
  \bibinfo{journal}{Phys. Rev. A} \textbf{\bibinfo{volume}{97}},
  \bibinfo{pages}{022319} (\bibinfo{year}{2018}).

\bibitem[{\citenamefont{Passarelli
  et~al.}(2020{\natexlab{a}})\citenamefont{Passarelli, Yip, Lidar, Nishimori,
  and Lucignano}}]{Passarelli_PRA20}
\bibinfo{author}{\bibfnamefont{G.}~\bibnamefont{Passarelli}},
  \bibinfo{author}{\bibfnamefont{K.-W.} \bibnamefont{Yip}},
  \bibinfo{author}{\bibfnamefont{D.~A.} \bibnamefont{Lidar}},
  \bibinfo{author}{\bibfnamefont{H.}~\bibnamefont{Nishimori}},
  \bibnamefont{and}
  \bibinfo{author}{\bibfnamefont{P.}~\bibnamefont{Lucignano}},
  \bibinfo{journal}{Phys. Rev. A} \textbf{\bibinfo{volume}{101}},
  \bibinfo{pages}{022331} (\bibinfo{year}{2020}{\natexlab{a}}).

\bibitem[{\citenamefont{Passarelli
  et~al.}(2020{\natexlab{b}})\citenamefont{Passarelli, Cataudella, Fazio, and
  Lucignano}}]{Passarelli_PRR20}
\bibinfo{author}{\bibfnamefont{G.}~\bibnamefont{Passarelli}},
  \bibinfo{author}{\bibfnamefont{V.}~\bibnamefont{Cataudella}},
  \bibinfo{author}{\bibfnamefont{R.}~\bibnamefont{Fazio}}, \bibnamefont{and}
  \bibinfo{author}{\bibfnamefont{P.}~\bibnamefont{Lucignano}},
  \bibinfo{journal}{Phys. Rev. Research} \textbf{\bibinfo{volume}{2}},
  \bibinfo{pages}{013283} (\bibinfo{year}{2020}{\natexlab{b}}).

\bibitem[{\citenamefont{Peruzzo et~al.}(2014)\citenamefont{Peruzzo, McClean,
  Shadbolt, Yung, Zhou, Love, Aspuru-Guzik, and O’brien}}]{Peruzzo_NatComm14}
\bibinfo{author}{\bibfnamefont{A.}~\bibnamefont{Peruzzo}},
  \bibinfo{author}{\bibfnamefont{J.}~\bibnamefont{McClean}},
  \bibinfo{author}{\bibfnamefont{P.}~\bibnamefont{Shadbolt}},
  \bibinfo{author}{\bibfnamefont{M.-H.} \bibnamefont{Yung}},
  \bibinfo{author}{\bibfnamefont{X.-Q.} \bibnamefont{Zhou}},
  \bibinfo{author}{\bibfnamefont{P.~J.} \bibnamefont{Love}},
  \bibinfo{author}{\bibfnamefont{A.}~\bibnamefont{Aspuru-Guzik}},
  \bibnamefont{and} \bibinfo{author}{\bibfnamefont{J.~L.}
  \bibnamefont{O’brien}}, \bibinfo{journal}{Nature communications}
  \textbf{\bibinfo{volume}{5}}, \bibinfo{pages}{4213} (\bibinfo{year}{2014}).

\bibitem[{\citenamefont{{Farhi} et~al.}(2014)\citenamefont{{Farhi},
  {Goldstone}, and {Gutmann}}}]{Farhi_arXiv2014}
\bibinfo{author}{\bibfnamefont{E.}~\bibnamefont{{Farhi}}},
  \bibinfo{author}{\bibfnamefont{J.}~\bibnamefont{{Goldstone}}},
  \bibnamefont{and}
  \bibinfo{author}{\bibfnamefont{S.}~\bibnamefont{{Gutmann}}},
  \bibinfo{journal}{arXiv e-prints} \bibinfo{eid}{arXiv:1411.4028}
  (\bibinfo{year}{2014}), \eprint{1411.4028}.

\bibitem[{\citenamefont{Kokail et~al.}(2019)\citenamefont{Kokail, Maier, van
  Bijnen, Brydges, Joshi, Jurcevic, Muschik, Silvi, Blatt, Roos
  et~al.}}]{Zoller_NAT19}
\bibinfo{author}{\bibfnamefont{C.}~\bibnamefont{Kokail}},
  \bibinfo{author}{\bibfnamefont{C.}~\bibnamefont{Maier}},
  \bibinfo{author}{\bibfnamefont{R.}~\bibnamefont{van Bijnen}},
  \bibinfo{author}{\bibfnamefont{T.}~\bibnamefont{Brydges}},
  \bibinfo{author}{\bibfnamefont{M.}~\bibnamefont{Joshi}},
  \bibinfo{author}{\bibfnamefont{P.}~\bibnamefont{Jurcevic}},
  \bibinfo{author}{\bibfnamefont{C.}~\bibnamefont{Muschik}},
  \bibinfo{author}{\bibfnamefont{P.}~\bibnamefont{Silvi}},
  \bibinfo{author}{\bibfnamefont{R.}~\bibnamefont{Blatt}},
  \bibinfo{author}{\bibfnamefont{C.}~\bibnamefont{Roos}}, \bibnamefont{et~al.},
  \bibinfo{journal}{Nature} \textbf{\bibinfo{volume}{569}},
  \bibinfo{pages}{355} (\bibinfo{year}{2019}).

\bibitem[{\citenamefont{McClean et~al.}(2016)\citenamefont{McClean, Romero,
  Babbush, and Aspuru-Guzik}}]{McClean_NewJPhys2016}
\bibinfo{author}{\bibfnamefont{J.~R.} \bibnamefont{McClean}},
  \bibinfo{author}{\bibfnamefont{J.}~\bibnamefont{Romero}},
  \bibinfo{author}{\bibfnamefont{R.}~\bibnamefont{Babbush}}, \bibnamefont{and}
  \bibinfo{author}{\bibfnamefont{A.}~\bibnamefont{Aspuru-Guzik}},
  \bibinfo{journal}{New Journal of Physics} \textbf{\bibinfo{volume}{18}},
  \bibinfo{pages}{023023} (\bibinfo{year}{2016}).

\bibitem[{\citenamefont{Niu et~al.}(2019)\citenamefont{Niu, Lu, and
  Chuang}}]{Niu_arx2019}
\bibinfo{author}{\bibfnamefont{M.~Y.} \bibnamefont{Niu}},
  \bibinfo{author}{\bibfnamefont{S.}~\bibnamefont{Lu}}, \bibnamefont{and}
  \bibinfo{author}{\bibfnamefont{I.~L.} \bibnamefont{Chuang}},
  \bibinfo{journal}{arXive e-prints}  (\bibinfo{year}{2019}),
  \eprint{1905.12134}.

\bibitem[{\citenamefont{Zhou et~al.}(2020)\citenamefont{Zhou, Wang, Choi,
  Pichler, and Lukin}}]{Zhou_PRX2020}
\bibinfo{author}{\bibfnamefont{L.}~\bibnamefont{Zhou}},
  \bibinfo{author}{\bibfnamefont{S.-T.} \bibnamefont{Wang}},
  \bibinfo{author}{\bibfnamefont{S.}~\bibnamefont{Choi}},
  \bibinfo{author}{\bibfnamefont{H.}~\bibnamefont{Pichler}}, \bibnamefont{and}
  \bibinfo{author}{\bibfnamefont{M.~D.} \bibnamefont{Lukin}},
  \bibinfo{journal}{Phys. Rev. X} \textbf{\bibinfo{volume}{10}},
  \bibinfo{pages}{021067} (\bibinfo{year}{2020}).

\bibitem[{\citenamefont{{Lloyd}}(2018)}]{Lloyd_arXiv2018}
\bibinfo{author}{\bibfnamefont{S.}~\bibnamefont{{Lloyd}}},
  \bibinfo{journal}{arXiv e-prints} \bibinfo{eid}{arXiv:1812.11075}
  (\bibinfo{year}{2018}), \eprint{1812.11075}.

\bibitem[{\citenamefont{Wang et~al.}(2018)\citenamefont{Wang, Hadfield, Jiang,
  and Rieffel}}]{Wang_PRA2018}
\bibinfo{author}{\bibfnamefont{Z.}~\bibnamefont{Wang}},
  \bibinfo{author}{\bibfnamefont{S.}~\bibnamefont{Hadfield}},
  \bibinfo{author}{\bibfnamefont{Z.}~\bibnamefont{Jiang}}, \bibnamefont{and}
  \bibinfo{author}{\bibfnamefont{E.~G.} \bibnamefont{Rieffel}},
  \bibinfo{journal}{Phys. Rev. A} \textbf{\bibinfo{volume}{97}},
  \bibinfo{pages}{022304} (\bibinfo{year}{2018}).

\bibitem[{\citenamefont{{Mbeng, Glen B. and Fazio, Rosario and Santoro,
  Giuseppe E.}}(2019)}]{Mbeng_arx19}
\bibinfo{author}{\bibnamefont{{Mbeng, Glen B. and Fazio, Rosario and Santoro,
  Giuseppe E.}}}, \bibinfo{journal}{arXiv e-prints}
  \bibinfo{eid}{arXiv:1906.08948} (\bibinfo{year}{2019}), \eprint{1906.08948}.

\bibitem[{\citenamefont{Jiang et~al.}(2017)\citenamefont{Jiang, Rieffel, and
  Wang}}]{Zhang_PRA2017}
\bibinfo{author}{\bibfnamefont{Z.}~\bibnamefont{Jiang}},
  \bibinfo{author}{\bibfnamefont{E.~G.} \bibnamefont{Rieffel}},
  \bibnamefont{and} \bibinfo{author}{\bibfnamefont{Z.}~\bibnamefont{Wang}},
  \bibinfo{journal}{Phys. Rev. A} \textbf{\bibinfo{volume}{95}},
  \bibinfo{pages}{062317} (\bibinfo{year}{2017}).

\bibitem[{\citenamefont{Mbeng et~al.}(2019{\natexlab{a}})\citenamefont{Mbeng,
  Fazio, and Santoro}}]{Mbeng_arXiv2019_digitizedQC}
\bibinfo{author}{\bibfnamefont{G.~B.} \bibnamefont{Mbeng}},
  \bibinfo{author}{\bibfnamefont{R.}~\bibnamefont{Fazio}}, \bibnamefont{and}
  \bibinfo{author}{\bibfnamefont{G.~E.} \bibnamefont{Santoro}},
  \emph{\bibinfo{title}{Optimal quantum control with digitized quantum
  annealing}} (\bibinfo{year}{2019}{\natexlab{a}}), \eprint{1911.12259}.

\bibitem[{\citenamefont{{Pagano} et~al.}(2019)\citenamefont{{Pagano}, {Bapat},
  {Becker}, {Collins}, {De}, {Hess}, {Kaplan}, {Kyprianidis}, {Tan}, {Baldwin}
  et~al.}}]{Pagano_arXiv2019}
\bibinfo{author}{\bibfnamefont{G.}~\bibnamefont{{Pagano}}},
  \bibinfo{author}{\bibfnamefont{A.}~\bibnamefont{{Bapat}}},
  \bibinfo{author}{\bibfnamefont{P.}~\bibnamefont{{Becker}}},
  \bibinfo{author}{\bibfnamefont{K.~S.} \bibnamefont{{Collins}}},
  \bibinfo{author}{\bibfnamefont{A.}~\bibnamefont{{De}}},
  \bibinfo{author}{\bibfnamefont{P.~W.} \bibnamefont{{Hess}}},
  \bibinfo{author}{\bibfnamefont{H.~B.} \bibnamefont{{Kaplan}}},
  \bibinfo{author}{\bibfnamefont{A.}~\bibnamefont{{Kyprianidis}}},
  \bibinfo{author}{\bibfnamefont{W.~L.} \bibnamefont{{Tan}}},
  \bibinfo{author}{\bibfnamefont{C.}~\bibnamefont{{Baldwin}}},
  \bibnamefont{et~al.}, \bibinfo{journal}{arXiv e-prints}
  \bibinfo{eid}{arXiv:1906.02700} (\bibinfo{year}{2019}), \eprint{1906.02700}.

\bibitem[{\citenamefont{Filippone et~al.}(2011)\citenamefont{Filippone, Dusuel,
  and Vidal}}]{Vidal_PRA11}
\bibinfo{author}{\bibfnamefont{M.}~\bibnamefont{Filippone}},
  \bibinfo{author}{\bibfnamefont{S.}~\bibnamefont{Dusuel}}, \bibnamefont{and}
  \bibinfo{author}{\bibfnamefont{J.}~\bibnamefont{Vidal}},
  \bibinfo{journal}{Phys. Rev. A} \textbf{\bibinfo{volume}{83}},
  \bibinfo{pages}{022327} (\bibinfo{year}{2011}).

\bibitem[{\citenamefont{Caneva et~al.}(2007)\citenamefont{Caneva, Fazio, and
  Santoro}}]{Caneva_PRB07}
\bibinfo{author}{\bibfnamefont{T.}~\bibnamefont{Caneva}},
  \bibinfo{author}{\bibfnamefont{R.}~\bibnamefont{Fazio}}, \bibnamefont{and}
  \bibinfo{author}{\bibfnamefont{G.~E.} \bibnamefont{Santoro}},
  \bibinfo{journal}{Phys. Rev. B} \textbf{\bibinfo{volume}{76}},
  \bibinfo{pages}{144427} (\bibinfo{year}{2007}).

\bibitem[{\citenamefont{Barends et~al.}(2016)\citenamefont{Barends, Shabani,
  Lamata, Kelly, Mezzacapo, Heras, Babbush, Fowler, Campbell, Chen
  et~al.}}]{Martinis_Nat16}
\bibinfo{author}{\bibfnamefont{R.}~\bibnamefont{Barends}},
  \bibinfo{author}{\bibfnamefont{A.}~\bibnamefont{Shabani}},
  \bibinfo{author}{\bibfnamefont{L.}~\bibnamefont{Lamata}},
  \bibinfo{author}{\bibfnamefont{J.}~\bibnamefont{Kelly}},
  \bibinfo{author}{\bibfnamefont{A.}~\bibnamefont{Mezzacapo}},
  \bibinfo{author}{\bibfnamefont{U.~L.} \bibnamefont{Heras}},
  \bibinfo{author}{\bibfnamefont{R.}~\bibnamefont{Babbush}},
  \bibinfo{author}{\bibfnamefont{A.~G.} \bibnamefont{Fowler}},
  \bibinfo{author}{\bibfnamefont{B.}~\bibnamefont{Campbell}},
  \bibinfo{author}{\bibfnamefont{Y.}~\bibnamefont{Chen}}, \bibnamefont{et~al.},
  \bibinfo{journal}{Nature} \textbf{\bibinfo{volume}{534}}, \bibinfo{pages}{222
  EP } (\bibinfo{year}{2016}).

\bibitem[{\citenamefont{Mbeng et~al.}(2019{\natexlab{b}})\citenamefont{Mbeng,
  Arceci, and Santoro}}]{Mbeng_dQA_PRB2019}
\bibinfo{author}{\bibfnamefont{G.~B.} \bibnamefont{Mbeng}},
  \bibinfo{author}{\bibfnamefont{L.}~\bibnamefont{Arceci}}, \bibnamefont{and}
  \bibinfo{author}{\bibfnamefont{G.~E.} \bibnamefont{Santoro}},
  \bibinfo{journal}{Phys. Rev. B} \textbf{\bibinfo{volume}{100}},
  \bibinfo{pages}{224201} (\bibinfo{year}{2019}{\natexlab{b}}).

\bibitem[{\citenamefont{Ho et~al.}(2019)\citenamefont{Ho, Jonay, and
  Hsieh}}]{Ho_PRA19}
\bibinfo{author}{\bibfnamefont{W.~W.} \bibnamefont{Ho}},
  \bibinfo{author}{\bibfnamefont{C.}~\bibnamefont{Jonay}}, \bibnamefont{and}
  \bibinfo{author}{\bibfnamefont{T.~H.} \bibnamefont{Hsieh}},
  \bibinfo{journal}{Phys. Rev. A} \textbf{\bibinfo{volume}{99}},
  \bibinfo{pages}{052332} (\bibinfo{year}{2019}).

\bibitem[{\citenamefont{Lipkin et~al.}(1965)\citenamefont{Lipkin, Meshkov, and
  Glick}}]{Lipkin_NucPhys65}
\bibinfo{author}{\bibfnamefont{H.}~\bibnamefont{Lipkin}},
  \bibinfo{author}{\bibfnamefont{N.}~\bibnamefont{Meshkov}}, \bibnamefont{and}
  \bibinfo{author}{\bibfnamefont{A.}~\bibnamefont{Glick}},
  \bibinfo{journal}{Nuclear Physics} \textbf{\bibinfo{volume}{62}},
  \bibinfo{pages}{188} (\bibinfo{year}{1965}).

\bibitem[{\citenamefont{Streif and Leib}(2019)}]{Streif_arXiv19}
\bibinfo{author}{\bibfnamefont{M.}~\bibnamefont{Streif}} \bibnamefont{and}
  \bibinfo{author}{\bibfnamefont{M.}~\bibnamefont{Leib}},
  \bibinfo{journal}{arXiv e-prints} \bibinfo{eid}{arXiv:1901.01903}
  (\bibinfo{year}{2019}), \eprint{1901.01903}.

\bibitem[{\citenamefont{Defenu et~al.}(2018)\citenamefont{Defenu, Enss,
  Kastner, and Morigi}}]{Defenu_PRL18}
\bibinfo{author}{\bibfnamefont{N.}~\bibnamefont{Defenu}},
  \bibinfo{author}{\bibfnamefont{T.}~\bibnamefont{Enss}},
  \bibinfo{author}{\bibfnamefont{M.}~\bibnamefont{Kastner}}, \bibnamefont{and}
  \bibinfo{author}{\bibfnamefont{G.}~\bibnamefont{Morigi}},
  \bibinfo{journal}{Phys. Rev. Lett.} \textbf{\bibinfo{volume}{121}},
  \bibinfo{pages}{240403} (\bibinfo{year}{2018}).

\bibitem[{\citenamefont{Nocedal and Wright}(2006)}]{Nocedal_book2006}
\bibinfo{author}{\bibfnamefont{J.}~\bibnamefont{Nocedal}} \bibnamefont{and}
  \bibinfo{author}{\bibfnamefont{S.}~\bibnamefont{Wright}},
  \emph{\bibinfo{title}{Numerical optimization}} (\bibinfo{publisher}{Springer
  Science \& Business Media}, \bibinfo{year}{2006}).

\bibitem[{\citenamefont{Wauters et~al.}(2020)\citenamefont{Wauters, Panizon,
  Mbeng, and Santoro}}]{Wauters_RL_arx2020}
\bibinfo{author}{\bibfnamefont{M.~M.} \bibnamefont{Wauters}},
  \bibinfo{author}{\bibfnamefont{E.}~\bibnamefont{Panizon}},
  \bibinfo{author}{\bibfnamefont{G.~B.} \bibnamefont{Mbeng}}, \bibnamefont{and}
  \bibinfo{author}{\bibfnamefont{G.~E.} \bibnamefont{Santoro}},
  \bibinfo{journal}{arXive e-prints}  (\bibinfo{year}{2020}),
  \eprint{2004.12323}.

\bibitem[{\citenamefont{Sutton and Barto}(2018)}]{Sutton_2ed}
\bibinfo{author}{\bibfnamefont{R.~S.} \bibnamefont{Sutton}} \bibnamefont{and}
  \bibinfo{author}{\bibfnamefont{A.~G.} \bibnamefont{Barto}},
  \emph{\bibinfo{title}{Reinforcement Learning, An Introduction}}
  (\bibinfo{publisher}{The MIT Press}, \bibinfo{year}{2018}),
  \bibinfo{edition}{2nd} ed.

\bibitem[{\citenamefont{{Ho} et~al.}(2018)\citenamefont{{Ho}, {Jonay}, and
  {Hsieh}}}]{Ho_arXiv2018}
\bibinfo{author}{\bibfnamefont{W.~W.} \bibnamefont{{Ho}}},
  \bibinfo{author}{\bibfnamefont{C.}~\bibnamefont{{Jonay}}}, \bibnamefont{and}
  \bibinfo{author}{\bibfnamefont{T.~H.} \bibnamefont{{Hsieh}}},
  \bibinfo{journal}{arXiv e-prints} \bibinfo{eid}{arXiv:1810.04817}
  (\bibinfo{year}{2018}), \eprint{1810.04817}.

\bibitem[{\citenamefont{Gauss}(1966)}]{book:Gauss1966}
\bibinfo{author}{\bibfnamefont{J.~C.~F.} \bibnamefont{Gauss}},
  \emph{\bibinfo{title}{Disquisitiones arithmeticae}}, vol.
  \bibinfo{volume}{157} (\bibinfo{publisher}{Yale University Press},
  \bibinfo{year}{1966}).

\bibitem[{\citenamefont{{Wikipedia contributors}}(2019)}]{wiki:mult_groups}
\bibinfo{author}{\bibnamefont{{Wikipedia contributors}}},
  \emph{\bibinfo{title}{Multiplicative group of integers modulo n ---
  {Wikipedia}{,} the free encyclopedia}} (\bibinfo{year}{2019}).

\end{thebibliography}

\end{document}